\documentclass[superscriptaddress,secnumarabic,
amssymb,amsmath,nobibnotes,aps,prd,showkeys,showpacs,nofootinbib]{revtex4-2}%
\usepackage{graphicx}
\usepackage{epsf}
\usepackage{bm}
\usepackage{amsmath}
\usepackage{amsfonts}
\usepackage{amssymb}
\usepackage{epstopdf}
\usepackage{subfigure}
\usepackage{natbib}
\usepackage{color}%
\usepackage{hyperref}
\hypersetup{dvips,dvipdfm,linktoc=page,colorlinks=true,linkcolor=blue,citecolor=red,filecolor=magenta,urlcolor=magenta,bookmarks=true}
\providecommand{\U}[1]{\protect\rule{.1in}{.1in}}
\newcommand{\be}{\begin{equation}}
	\newcommand{\ee}{\end{equation}}

\newcommand{\mincir}{\raise
	-3.truept\hbox{\rlap{\hbox{$\sim$}}\raise4.truept\hbox{$<$}\ }}
\newcommand{\magcir}{\raise
	-3.truept\hbox{\rlap{\hbox{$\sim$}}\raise4.truept\hbox{$>$}\ }}

\begin{document}
\title{Interacting tachyon with varying mass dark matter} 
\author{Goutam Mandal}
\email{gmandal243@gmail.com; 
	rs\_goutamm@nbu.ac.in}
\affiliation{Department of Mathematics, University of North Bengal, Raja Rammohunpur, Darjeeling-734013, West Bengal, India.}
\author{Sudip Mishra}
\email{sudipcmiiitmath@gmail.com}
\affiliation{Department of Applied Mathematics, Maulana Abul Kalam Azad University of Technology, Haringhata, Nadia- 741249, West Bengal, India.}
\author{Abdulla Al Mamon}
\email{abdulla.physics@gmail.com}
\affiliation{Department of Physics, Vivekananda Satavarshiki Mahavidyalaya (affiliated to the Vidyasagar University), Manikpara 721513, West Bengal, India.}
\author{Sujay Kr. Biswas\footnote{corresponding author}}
\email{sujaymathju@gmail.com; sujay.math@nbu.ac.in}
\affiliation{Department of Mathematics, University of North Bengal, Raja Rammohunpur, Darjeeling-734013, West Bengal, India.}
\begin{abstract}
This paper presents an investigation of cosmological dynamics of tachyon fluid coupled to varying-mass dark matter particles in the background of spatially flat FLRW universe. The mechanism of varying mass particles scenario assumes the mass of the dark matter depends on time `$t$' through the scalar field $\phi$ in the sense that the decaying of dark matter reproduces the scalar field. First, we analyze the model from dynamical systems perspective by converting the cosmological evolution equations into an autonomous system of ordinary differential equations with a suitable transformation of variables. We choose the mass of dark matter as exponential function of scalar field and the exponential potential of the tachyon field is undertaken in such a way that the autonomous system is reduced in three dimensional form. The critical points obtained from the system are non-hyperbolic in nature. The center manifold theory is employed to discuss the nature of the critical points. Numerical investigation also carried out for some critical points. From this analysis, we obtain dust dominated decelerated transient phase of the universe followed by dark energy dominated scaling attractor alleviating the coincidence problem. Next, we perform the statefinder diagnostic approach to compare our model to $\Lambda$CDM and finally we study the evolution of the Hubble parameter and the distance modulus and compare this with observational data.
\end{abstract}
\maketitle

\section{Introduction}

The fact that the universe is at the current epoch undergoing an accelerated expansion phase has been well established by various independent observational data \cite{Riess:1998cb,Perlmutter:1998np,Betoule:2014frx,Ade:2013zuv,Ade:2015xua}. One of the most significant challenges in modern physics is explaining the fundamental cause of this observed acceleration. Theoretically, the matter or gravitational sector requires to be modified in order to provide a proper explanation for accelerating universe. This led to the modification of the matter source with dark matter (DM) and dark energy (DE). DE is usually characterized by a equation of state (EoS) parameter, $\omega_{\Lambda}=-1$.
The prevailing belief is that DE poses huge negative pressure (repulsive gravity) and dominates over DM component at present. Several DE models have been investigated in the literature and one can refer to \cite{E. Copeland2006,bambader} for review. Among them, cosmological constant $\Lambda$ model is the simplest one, which along with cold dark matter (CDM) commonly referred to as the standard $\Lambda$CDM model, shows the best fit according the several observational data. However, cosmological constant has some theoretical issues like `fine tuning' \cite{S. Weinberg1989, V. Sahni2000, T. Padmanabhan2003}  and `coincidence' \cite{I. Zlatev1999} (why the energy density associated with cosmological constant is very small when expressed in natural units and since it could have had two distinct contributions from the  matter and gravitational parts, why does the sum remain so fine tuned $?$) problems. In an attempt to address these problems, many ideas involving time-varying DE based on scalar field have been proposed such as quintessence, k-essence, phantom, tachyon and many more \cite{E. Copeland2006,bambader}.\\

Inflation provides a technique for production of density perturbations which is required to mould the evolution of the universe.  In the simple case of inflation, the universe is dominated with a scalar field and potential energy dominates over the kinetic term, followed by a reheating period \cite{Gibbons2003, Jassal2004}.  But there is no well accepted exposition to integrate inflationary scenario,  the scalar field that drives inflation,  with one of the known fields of particle physics.   It is also salient to emerge inflation potential naturally from underlying fundamental theory.  In this regard, tachyon fields analogous to unstable D-branes could be accountable for inflation in early time \cite{Sami2004, Nozari_Rashidi2013}.
It has been first shown in Ref.\cite{Mazumdar2001} that the mechanism of decaying of non-BPS D4-brane into a stable D3 brane will give rise to a tachyon field via tachyon condensation, and that can act as a cosmological inflating field. After that, Sen in Ref.\cite{Sen2002a,Sen2002b,Sen2002c} showed that rolling tachyon field can arise naturally from unstable D-branes in bosonic and superstring theories. Further, he showed in Ref.\cite{Sen2002c} that the effective field of such tachyon may be designed as Dirac-Born-Infeld (DBI) scalar field theory. 
Furthermore, investigations of DBI field have been carried out in Refs.\cite{Padmanabhan2002,Gibbons2002} in the framework of FLRW cosmology to give inflating field behaviour. \\

It should be mentioned that even in bosonic string theory, the quantization of the Polyakov action (conformal transformation of Nambu-Goto action) provides a tachyonic like field giving locally unstable nature due to reverse sign of kinetic term, however, which soon decays through spontaneous symmetry breaking \cite{Green1988,Polchinski2005}. In the present work, we have undertaken DBI field as effective field theory D-branes, so, there are no more instability problems as all the instable modes have been integrated out as shown in references \cite{Sen2002b,Sen2002c}.

 In order to address such issues, many theoretical models of DE  have been raised in the theoretical background among them interacting cosmological models where DE interacts with DM have gained a remarkable attention with a successive number of observational data \cite{E. Copeland2006}. Infact, latest observations indicate that there could be an non-vanishing interaction within $1\sigma$ confidence region in dark sectors \cite{int1, nps,kn}. It should also be mentioned that the authors in Ref. \cite{Andre A.Costa2014} have tested a phenomenological interacting model with Planck data. After that another interacting model in Ref.\cite{J.S.Wang2015} is tested by using SNeIa, BAO measurements, Hubble parameter data and CMB observations. Further, an interaction term with linear combination of energy densities of dark matter and dark energy was investigated in Ref. \cite{S.Pan2015mnras} with the SNeIa data. It was shown that for a small value of the coupling of interaction, the model tends to $\Lambda$CDM.

Due to unknown nature of two dark components there is no standard form of interaction term even there is no any guiding principle to choose an interaction between them. Therefore, one can adopt the interaction  purely on the basis of phenomenology. In fact, an appropriate interaction can provide a good mechanism to alleviate the coincidence problem. The models of DE interacting with DM have extensively been studied in the literature such as an interacting quintessence is investigated in Ref.\cite{C.G.Bohmer2008}  where the interaction terms are taken in terms of energy densities and other covariant quantities. A detailed analysis of interacting quintessence is performed at the perturbative level (see in Ref. \cite{N.Tamanini2015}). Linear and non-linear interactions have been investigated in context of phantom cosmology \cite{Xi-ming Chen2009}. A class of linear and non-linear interactions have been investigated in Refs. \cite{L.P.Chimento2010,L.P.Chimento2012}  where the interaction terms depend on energy densities of dark energy, dark matter and their derivatives.
Interacting Chaplygin gas with different linear and non-linear sign changeable interactions are studied in Ref.\cite{M.Khurshudyan2015}. For a detailed review on linear and non-linear interactions, one can follow the review paper \cite{Yuri.L.Bolotin2014}. Further, cosmological models interacting scalar field are investigated in modified gravity in Refs. \cite{S.Kr.Biswas2015a,S.Kr.Biswas2015b}. In Ref. \cite{T.Harko2013}, the authors have studied the interaction between dark energy-dark matter in the framework of irreversible thermodynamics of open system with particle creation. Interacting model of dark energy in Brans Dicke theory was investigated in Ref. \cite{sdaam2014} and interacting Barrow holographic dark energy is investigated in Ref. \cite{aamapss2021}. Recent studies (see in references \cite{Mandal2025a,Mandal2025b}) explore the dynamical systems analysis of interacting Umami chaplygin gas and interacting holographic dark energy in framework of adiabatic particle creation mechanism which show different cosmic phases of the universe. 
As nothing much is known about the physical nature of DE, probe is still on to hunt for an appropriate candidate for DE. \\

On the other hand, interacting DE with varying mass DM particles can also provide a proper way to alleviate coincidence problem. The model of varying mass DM particles is based on the mechanism that decaying of DM reproduces the scalar field as DE. Here, mass of the dark matter is dependent on cosmic time `$t$' via a scalar field `$\phi$'. It should be mentioned that this mechanism can be treated equivalent to an interaction included in the dark sectors, see for example \cite{Teixeira2019, Honorez2010}. However, the model of varying mass has some basic differences from interacting models. The mass of dark matter can be exponential or power-law functions of scalar field or any other general form. Interacting quintessence studied recently with varying mass dark matter in the Lyra's manifold  in context of dynamical analysis, where mass of dark matter is assumed to be vary a exponential function of scalar field \cite{Mandal2021}. Interacting phantom is also studied where exponential and power-law of scalar field as well as exponential or power-law mass dependence is undertaken in the perspective of dynamical system \cite{Leon2010}. A non-interacting scalar field model is investigated with arbitrary potential in Ref. \cite{Leon2023} Consequently, a center manifold theory is employed to study the interacting phantom in association with varying mass dark matter \cite{S.Chakraborty2020}. A plethora of interacting quintessence models have been put forward over the years in the literature \cite{Zhang2005,Comelli2003,Franca2004} where the dark matter mass depends on exponential or power-law of scalar field $\phi$, and these provided the possible solutions to coincidence problem. In this context, we would like to mention that the tachyon field can be made as a satisfactory candidate for the
high energy inflation \cite{Mazumdar2001} and at the same time as an origin of DE depending on the form of the tachyon potential \cite{Padmanabhan2002}. The DBI scalar field has been studied in context of modified symmetric teleparallel gravity in Ref. \cite{Ghosh2024}. Also, some recent literature that studies cosmological application can be explored (for details see \cite{Srivastava2025,Fayaz2013}).
 \\

Starting from the above premises, in this paper, we shall investigate more in-depth the dynamics of tachyon scalar field that is  interacting with varying mass dark matter particles where mass of dark matter varies as exponential function of scalar field, or a general form of scalar field in the dynamical systems perspective. From the dynamical system analysis, we obtain some physically meaningful solutions representing the early dark matter (specifically dust) dominated solution which describes the decelerated phase of the universe and late-time de Sitter solution also obtained representing the accelerated universe. Interestingly, the DE-DM scaling solutions are also obtained satisfying the same order of energy densities solving the coincidence problem.
Furthermore, we obtained the analytic expression of Hubble parameter for the model under consideration. We then study the evolution of different diagnostic parameter pairs for the derived model and compare that with the $\Lambda$CDM model. Finally, we study the evolution of the Hubble parameter and the distance modulus, and compare that with the observational Hubble parameter data and Type Ia Supernovae
data, respectively.\\

The paper is organized as follows. In the next section \ref{model and autonomous system}, we first discuss about the cosmological model considered here and then discuss about the formation of autonomous system and the relevant parameters. Phase space analysis has also been presented in section \ref{phase space autonomous system}. In addition, in section \ref{statefinder}, we examine the behavior of diagnostic pair parameters. In section \ref{data} we depicts the evolution of different cosmological
quantities for our model and compare it with the observational data. Finally, our main findings and conclusions are shortly discussed in section \ref{conclusion}. Throughout the paper, we use natural units in which $\kappa^{2}=8\pi G=c=1$.      


\section{ Model for interacting tachyon with varying mass dark matter particles and autonomous system}\label{model and autonomous system}

In this section we shall first discuss the model of tachyonic DE fluid coupled to a varying mass dark matter particles and then discuss about the formation of autonomous system from the cosmological evolution equations after suitable transformation of variables.

\subsection{The model of interacting varying mass tachyonic DE}

According to cosmological principle and from the survey of superclusters of galaxies \cite{Einasto2006}, it is evident that the universe's spatial section is homogeneous and isotropic at sufficiently large scales. Observation also predicts that the universe is spatially flat \cite{dH0}. Thus, we consider the background universe which is in good agreement with observations, described by maximally symmetric space-time that is well described by the following flat Friedmann-Lema\^itre-Robertson-Walker (FLRW) metric \cite{Hawking and Ellis 1973}:
\begin{equation}\label{FLRW metric}
	ds^2=-dt^2 +a^{2}(t) \left[ dr^2 +r^2 (d\theta^2 +sin^2 \theta d\phi^2)\right]
\end{equation}  
in which  $a(t)$ is the scale factor of the universe and $t$ is the cosmic time.  Under the above scenario, the Friedmann equations can be obtained as

\begin{equation}\label{Friedmann}
	3 H^{2} =\rho_{Total},
\end{equation}
\begin{equation}\label{Raychaudhuri}
	- 2 \dot{H} =\rho_{Total}+p_{Total}
\end{equation}
where $H=\frac{\dot{a}}{a}$ is Hubble parameter and an over `dot' stands for the differentiation with respect to the cosmic time $t$. Also, $\rho_{Total}$ and $p_{Total}$ are the effective (total) energy density and pressure of all matter content in the universe. We consider here pressureless ($p_{m}=0$) dust as DM and tachyonic fluid as DE as the main constituent of the universe. Therefore,  $\rho_{Total}$ and $p_{Total}$ are given by 
\begin{equation}\label{rho tot}
	\rho_{Total}=\rho_{m}+\rho_{\phi}
\end{equation}
and
\begin{equation}\label{p tot}
	p_{Total}=p_\phi
\end{equation}
where $\rho_{m}$ denotes the energy density of the DM. The energy density and the thermodynamic pressure for the tachyonic DE are denoted by $\rho_{\phi}$ and $p_{\phi}$ and are defined as follows \cite{Mazumdar2001,Sen2002a,Sen2002b,Sen2002c}: 
\begin{equation}\label{energy density scalar}
	\rho_{\phi}=\frac{V(\phi)}{\sqrt{1-\dot{\phi}^{2}}}
\end{equation}
and 
\begin{equation}\label{pressure scalar}
	p_{\phi}=-V(\phi)\sqrt{1-\dot{\phi}^{2}},
\end{equation}
where $V(\phi)$ is the potential function of the scalar field and $\dot{\phi}^2$ is the kinetic part of the scalar field. The EoS parameter for scalar field ($\omega_{\phi}$) and the total (effective) EoS parameter for all matter content of the universe ($\omega_{eff}$) are given by
\begin{eqnarray}
	\omega_{\phi}&=&\frac{p_{\phi}}{\rho_{\phi}},\\
	\omega_{eff}&=&\frac{p_{Total}}{\rho_{Total}}= \frac{p_{\phi}}{\rho_{m}+\rho_{\phi}}.\label{equations}
\end{eqnarray}
The energy conservation equation for total matter content of the universe will take the form 
\begin{equation}\label{continuity}
	\dot{\rho}_{Total}+3H(\rho_{Total}+p_{Total})=0
\end{equation}
From eqs. (\ref{Friedmann}) and (\ref{energy density scalar}), one can obtain the modified Friedmann equation as:
\begin{equation}\label{Friedmann 1}
	3H^{2}=\rho_{m}+\frac{V(\phi)}{\sqrt{1- \dot{\phi}^{2}}} 
\end{equation}
and from eqs. (\ref{Raychaudhuri}) and (\ref{pressure scalar}), one can write the acceleration equation as:
\begin{equation}
	-2 \dot{H}=\rho_{m}+\frac{V(\phi)}{\sqrt{1- \dot{\phi}^{2}}}-V(\phi)\sqrt{1- \dot{\phi}^{2}} 
\end{equation}
Now we consider tachyon scalar field (DE) interacts with varying mass DM particles where mass of DM varies with scalar field $\phi$. According to the variable mass particle scenario (VAMP), mass of dark matter particles depend on time `$t$' through scalar field $\phi$. Since cold dark matter (CDM) particles are stable, its number density must obey the following conservation equation:
\begin{equation}\label{number density}
	\dot{n}_{m}+3Hn_m=0
\end{equation}
where, $n_{m}$ is number density of DM particles. Here, we denote the mass of DM particles as $Q_m$ which is assumed to be dependent on scalar field $\phi$ and this leads to the fact that the energy density $\rho_{m}$ is also depended on $\phi$ by the following relation:
\begin{equation}\label{energy density number density}
	\rho_{m}(\phi)=Q_m (\phi) n_m
\end{equation}
Using eq. (\ref{number density}), the time-derivative of eq. (\ref{energy density number density}) will lead to the following modified conservation equation for DM
\begin{equation}\label{continuity DM}
	\dot{\rho}_{m}+3H\rho_m=\frac{Q_{m}^{'}(\phi)}{Q_{m}(\phi)}\dot{\phi}\rho_{m},
\end{equation}
where the prime  stands for derivative with respect to scalar field $\phi$ (i.e., $'\equiv \frac{d}{d\phi}$). By observing eqs. (\ref{continuity}) and (\ref{continuity DM}), we can obtain in a similar manner
\begin{equation}\label{continuity DE}
	\dot{\rho}_{\phi}+3H(1+\omega_{\phi})\rho_{\phi}=-\frac{Q_{m}^{'}(\phi)}{Q_{m}(\phi)}\dot{\phi}\rho_{m}=-Q
\end{equation}
This above equation is termed as conservation equation for DE in presence of varying mass DM particles. In this case, the term $\frac{Q_{m}^{'}(\phi)}{Q_{m}(\phi)}\dot{\phi}\rho_{m}$ in the right hand part of eqs. (\ref{continuity DM}) and (\ref{continuity DE}) plays the role of an interaction between the dark sectors (DE $\&$ DM) where $Q_{m}^{'}(\phi)\dot{\phi}<0$ indicates that there is an energy transfer from the DM to DE while $Q_{m}^{'}(\phi)\dot{\phi}>0$ refers to the fact that the energy flow occurs in the opposite direction. For the sake of simplicity, here we denote the interaction term by 
\begin{equation}\label{eq-intqbeta}
	Q=\frac{Q_{m}^{'}(\phi)}{Q_{m}(\phi)}\dot{\phi}\rho_{m}=\tilde{\beta}(\phi)\dot{\phi}\rho_{m}
\end{equation} 
where $\tilde{\beta}(\phi)=\frac{\partial}{\partial \phi}ln(Q_{m}(\phi))$. It should be noted that Honorez et al. \cite{Honorez2010} studied a coupled quintessence model in which the interaction with the dark matter sector is a function of the quintessence potential. They have also showed that such type of interaction (\ref{eq-intqbeta}) can arise from a field dependent mass term for the DM field. This kind of coupling (i.e., $\tilde{\beta}(\phi)=$ constant) can be found in the literature where non-minimally coupled Brans-Dicke theory containing a
self-interacting potential, can automatically give rise to an interaction
between the Brans-Dicke scalar field and the normal matter by applying a suitable conformal transformation (see \cite{sdaam2014} and the references therein).

Note that due to an interaction between the pressure-less DM and tachyon scalar field as DE, the evolution of the pressure-less DM does not follow the standard evolution $a^{-3}$. This is expected since an interaction will affect the evolution of the dark fluids.  However,  significant deviation from $\Lambda$CDM is not expected because if the coupling parameter is found to be large that may significantly affect the CMB spectrum. As $\Lambda$CDM is quite successful but still its revision is necessary according to its limitations from both observational and theoretical directions \cite{DiValentino:2021izs},  therefore, one may wonder that the interacting scenario could be an approximate version of the $\Lambda$CDM model. Moreover, observational data also indicate that the coupling parameter is small and hence, interacting scenarios can explain the evolution of the universe at the level of perturbations~\cite{Pan:2019,Yang:2018}.

In our case, it is also clearly seen that the tachyonic scalar field and the DM do not evolve independently  but interact with each other via an interaction term $Q$ in the energy conservation eqs. (\ref{continuity DM}) and (\ref{continuity DE}). It is important to note here that if we concentrate on the other interacting DE-DM models, the form of coupling chosen is ad-hoc and the source of such coupling is not known (see e.g.  \cite{Yuri.L.Bolotin2014}). But, in our case, unlike other models, the interaction term is not an input but obtained its form from the field equations by considering the variable mass particle scenarios as already referred earlier. This makes our work very interesting and deserves further study in the present context. \\

Moreover, one can also express the conservation eqs. (\ref{continuity DM}) and (\ref{continuity DE}) in non-interacting
form with effective EoS parameters for DM and DE as follows: 
\begin{eqnarray}
	\dot{\rho}_{m}+3H\rho_m(1+\omega^{(m)}_{eff})=0,\\ \label{eq-eeffrmceq}
	\dot{\rho_\phi}+3H\rho_{\phi}(1+\omega^{(\phi)}_{eff})=0, \label{eq-eeffrphiceq}
\end{eqnarray}
where $\omega^{(m)}_{eff}=-\frac{Q_{m}^{'}(\phi)}{Q_{m}(\phi)}\frac{\dot{\phi}}{3H}$ and $\omega^{(\phi)}_{eff}=\omega_{\phi}+\frac{Q_{m}^{'}(\phi)}{Q_{m}(\phi)}\frac{\dot{\phi}}{3H}\frac{\rho_{m}}{\rho_{\phi}}$ are the effective EoS parameter for DM and DE, respectively.
Finally, using eqs. (\ref{energy density scalar}), (\ref{pressure scalar}) and (\ref{continuity DE}), one can now derive the evolution equation for tachyon field $\phi$ in the VAMP scenario as 
\begin{equation}\label{EvolutionEqn-scalar}
	\frac{\ddot{\phi}}{1- \dot{\phi}^{2}}+\frac{V'}{V}+3H \dot{\phi}=-\frac{Q_{m}^{'}(\phi)}{Q_{m}(\phi)} \rho_{m} \frac{\sqrt{1-\dot{\phi}^{2}}}{V}
\end{equation}  
The dynamics of cosmological evolution of the universe in this scenario of VAMP, we will perform the critical points analysis in the next section by transforming the cosmological evolution equations into an autonomous system of first order ordinary differential equations \cite{Xi-ming Chen2009}:
\begin{equation}
\textbf{x}'=\textbf{f(x)}	
\end{equation}	
where `prime' denotes the derivative with respect to $N=\ln (a)$, and $\textbf{x}$ is the column vector made by the auxiliary (dynamical) variables, and $\textbf{f(x)}$ is also the column vector corresponding to the autonomous equations. We are to extract critical point $x=x_c$ by equating $\textbf{f(x)}=0$. To determine the stability of the critical point, we have to perturb the system around the critical point. The small perturbation is taken as $x=x_c +\xi$, where $\xi$ is a column vector, perturbation of variables. Then, the linearised system is
\begin{equation}
	\xi'=J \xi,	
\end{equation}
where $J$ is the Jacobian matrix for the transformation, containing the coefficients of perturbation equations. The eigenvalues of the linearised Jacobian matrix will determine the nature of critical point $x=x_c$.

\subsection{Autonomous system and cosmological parameters}\label{autonomous system and parameters}
In this section, we shall discuss the construction of the autonomous system by adopting suitable dynamical variables.
For qualitative analysis, we choose the following dimensionless variables 
\begin{equation}\label{dynamical_variables}
	x=\dot{\phi},~~~y^2=\frac{V}{3H^2},~~~z=\frac{H_{0}}{H+H_{0}},~~~\alpha=-\frac{1}{H_{0}}\frac{Q_{m}^{'}(\phi)}{Q_{m}(\phi)}~~~\mbox{and}~~~\beta=\frac{1}{H_{0}}\frac{V'}{V}
\end{equation}
which are normalized over Hubble scale. \\
With these variables (\ref{dynamical_variables}) the above cosmological equations can be converted into the following 5D autonomous system as
\begin{eqnarray}
	\begin{split}
		\frac{dx}{dN}& = \left(1-x^2\right) \left\lbrace -3 x-\frac{\beta  z}{1-z}+ \frac{\left(\sqrt{1-x^2}-y^2\right) \alpha  z}{y^2 (1-z)} \right\rbrace ,& \\
		\frac{dy}{dN}& = \frac{y}{2} \left\lbrace \frac{\beta  x z}{1-z}+3 \left(1-y^{2}\sqrt{1-x^2} \right)\right\rbrace ,& \\
		\frac{dz}{dN} & =\frac{3}{2} z (1-z) \left(1-y^{2}\sqrt{1-x^2} \right),&\\
		\frac{d\alpha}{dN} &=-\frac{\alpha^{2} xz f(\alpha)}{1-z},&\\
		\frac{d\beta}{dN} &=\frac{\beta^{2} xz g(\beta)}{1-z}
		&~~\label{autonomous_system 1}
	\end{split}
\end{eqnarray}
where $ N=\ln a $ is the e-folding parameter taken to be independent variable and $f(\alpha)=\frac{Q_{m}'' Q_{m}}{Q_{m}'^{2}}-1$ and $g(\beta)=\frac{V'' V}{V'^{2}}-1$.\\

\subsubsection{\bf{Autonomous system with exponential mass function and exponential potential}}

In the model of exponential variable mass particles of DM, we consider the mass of DM is function of scalar field $\phi$ as
$Q_{m}(\phi)=Q_{m0}~ \mbox{exp} \{-\alpha H_{0}\phi\}$ and the potential of the scalar field $\phi$ as $V(\phi)=V_{0}~exp\{\beta H_{0} \phi\}$, where $Q_{m0}$ and $V_{0}$ are constant and $\alpha$ and $\beta$ are constant parameters.
Now with the exponential potential and exponential mass function of DM, the 5D autonomous system (\ref{autonomous_system 1}) reduces to the following 3D autonomous system of ordinary differential equations:
\begin{eqnarray}
	\begin{split}
		\frac{dx}{dN}& = \left(1-x^2\right) \left\lbrace -3 x-\frac{\beta  z}{1-z}+ \frac{\left(\sqrt{1-x^2}-y^2\right) \alpha  z}{y^2 (1-z)} \right\rbrace ,& \\
		\frac{dy}{dN}& = \frac{y}{2} \left\lbrace \frac{\beta  x z}{1-z}+3 \left(1-y^{2}\sqrt{1-x^2} \right)\right\rbrace ,& \\
		\frac{dz}{dN} & =\frac{3}{2} z (1-z) \left(1-y^{2}\sqrt{1-x^2} \right),
		&~~\label{autonomous_system}
	\end{split}
\end{eqnarray}
where $ N=\ln a $ is the e-folding parameter taken to be independent variable.\\

It is clear that the autonomous system (\ref{autonomous_system}) has singularities at $y=0$ and $z=1$. In this situation we can not study the properties of solutions lying on $y=0$ and $z=1$ plane. In order to remove the singularities, we multiply the right hand sides of the system (\ref{autonomous_system}) by $y^{2}(1-z)$. This operation allows us to analyze the solutions on $y=0$ and $z=1$ plane without changing the qualitative dynamical features of the system in the other regions of the phase space for $y^{2}(1-z)>0$. After applying this technique, we obtain 

\begin{eqnarray}
	\begin{split}
		\frac{dx}{dN}& = \left(1-x^2\right) \left\lbrace -3 x y^{2}(1-z)-\beta y^{2} z+ \left(\sqrt{1-x^2}-y^2\right) \alpha  z\right\rbrace ,& \\
		\frac{dy}{dN}& = \frac{y^{3}}{2} \left\lbrace \beta  x z +3 (1-z) \left(1-y^{2}\sqrt{1-x^2} \right)\right\rbrace ,& \\
		\frac{dz}{dN} & =\frac{3}{2}y^{2} z (1-z)^{2} \left(1-y^{2}\sqrt{1-x^2} \right),
		&~~\label{autonomous_system 2}
	\end{split}
\end{eqnarray}
which is now regular at  $y=0$ and $z=1$ planes.

.\\


\subsubsection{\bf{cosmological parameters}}
\begin{figure}
		\centering
		\subfigure[]{
			\includegraphics[width=0.3\textwidth]{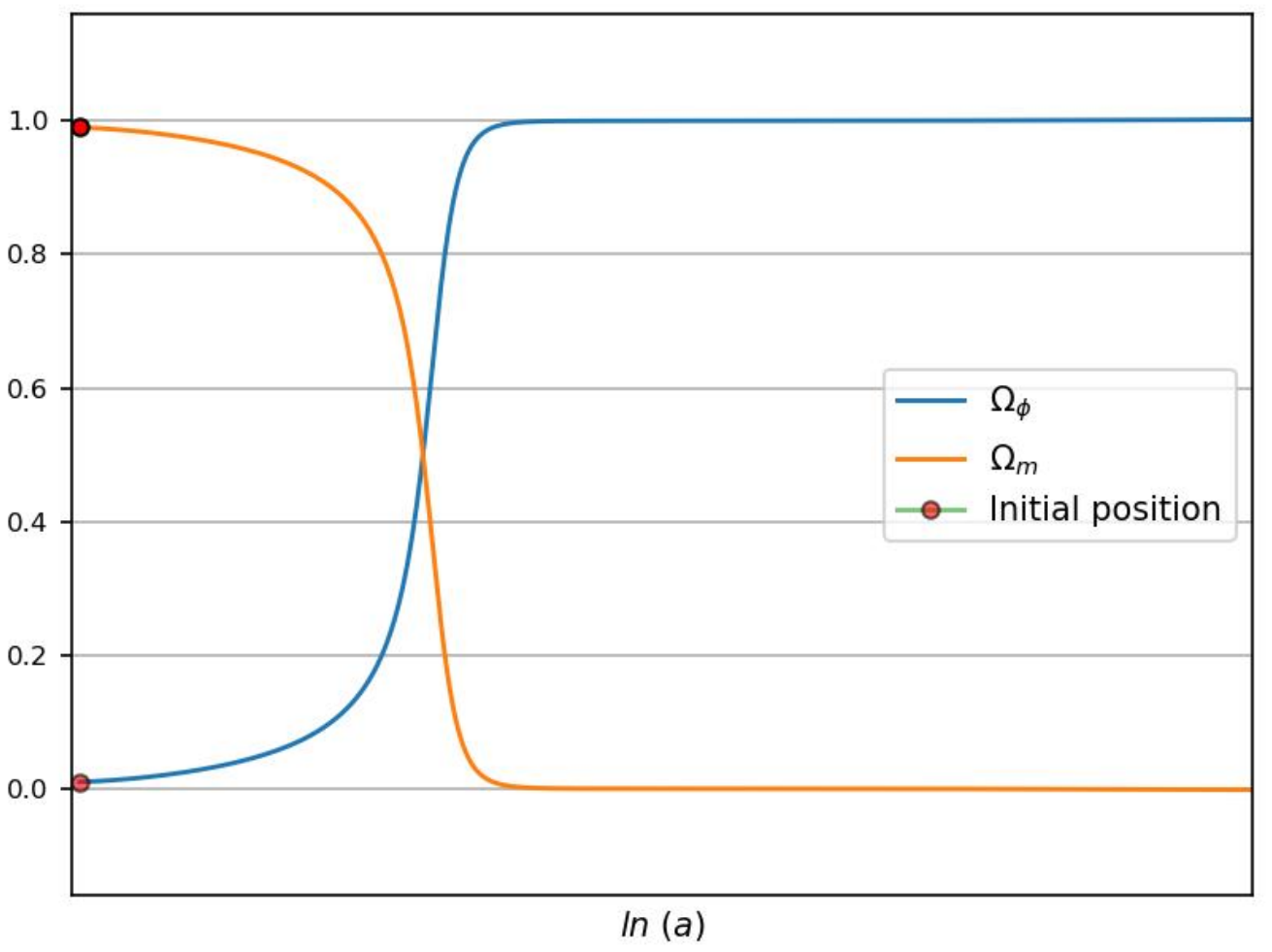}\label{test1}}
		\subfigure[]{
			\includegraphics[width=0.3\textwidth]{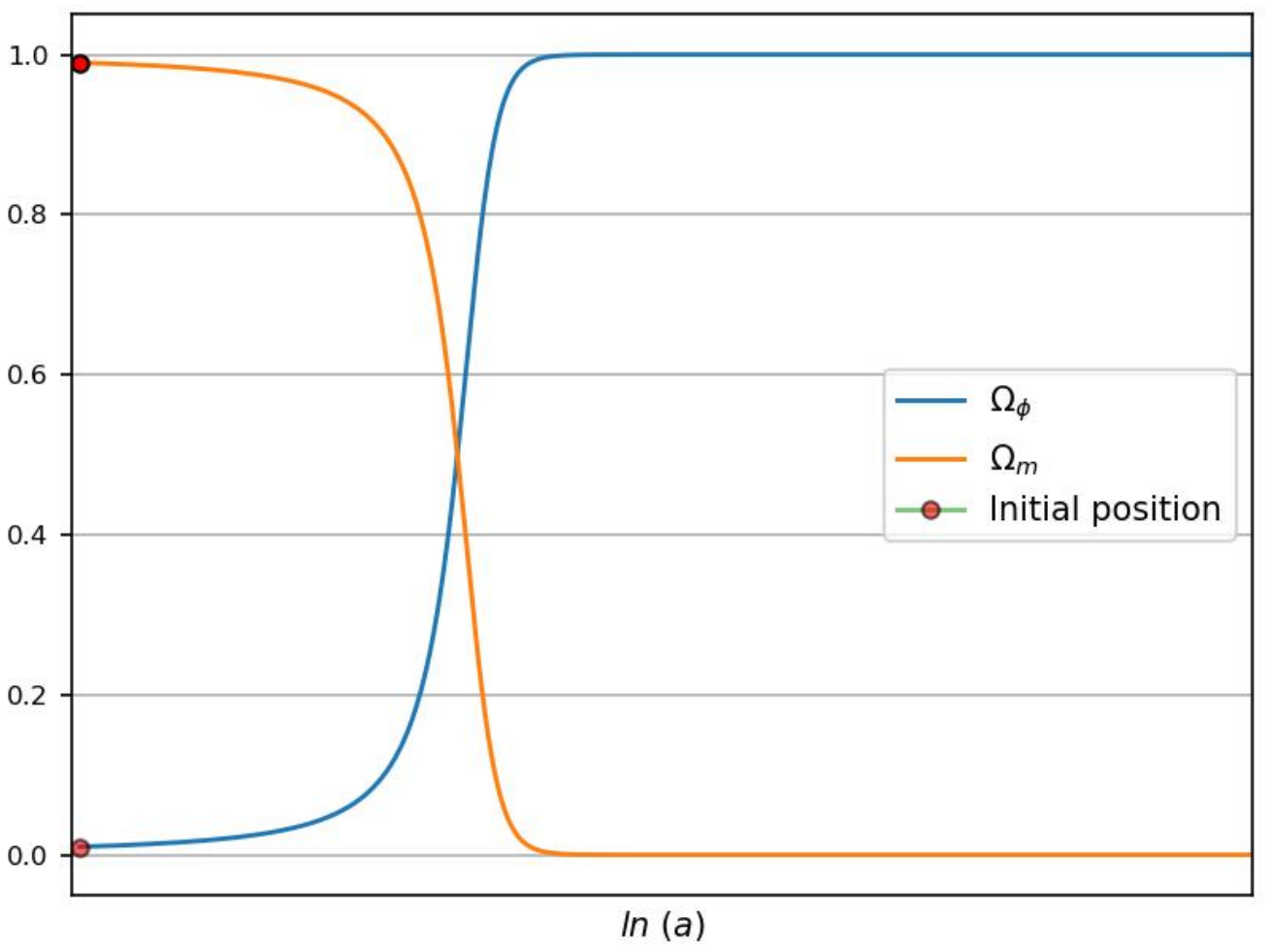}\label{test1a}}
		\subfigure[]{
			\includegraphics[width=0.3\textwidth]{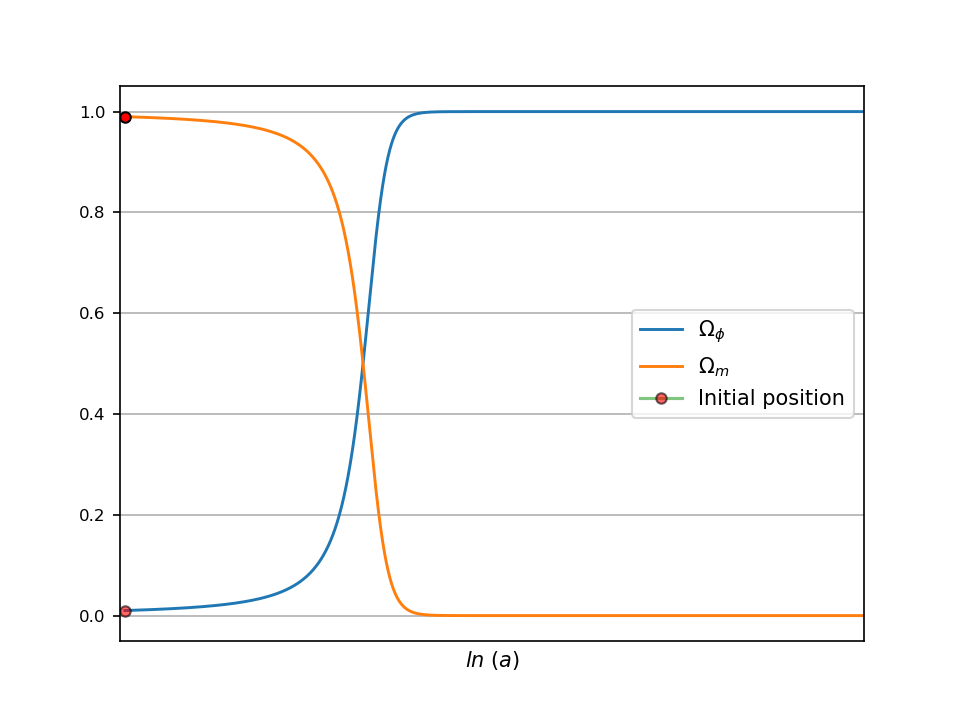}\label{test1b}}
		\caption{ The figures show the evolution of density parameters for tachyon and dark matter for different initial conditions.  Panel (a) for $\alpha=0.5,~\beta=0.5$. Panel (b) for $\alpha=0.001,~\beta=0.1$. Panel (c) for $\alpha=-0.001,~\beta=-0.1$.}
		\label{test}
	\end{figure}

Now we obtain the cosmological parameters in terms of dynamical variables as follows :\\
Density parameters for tachyon scalar field (DE) and dark matter are 
\begin{equation}\label{density_parameter}
	\Omega_{\phi}=\frac{y^{2}}{\sqrt{1-x^{2}}},
\end{equation}
and
\begin{equation}\label{density_parameter_m}
	\Omega_{m}=1-\frac{y^{2}}{\sqrt{1-x^{2}}},
\end{equation}
It should be noted that the fig.\ref{test} is plotted for different values of parameters ($\alpha$ and $\beta$) with different initial conditions, and it has been shown that the model does not exhibit any negative energy density during the evolution of the universe for a wide range of coupling. Thus, the present interacting tachyon model behaves well in cosmological point in view, where the interaction $Q$ ensures physically consistent energy transfer, such as consistent with conservation laws, avoiding negative energy density and instabilities in cosmic evolution. 

The effective equation of state parameter of tachyon field (DE) for this model read as
\begin{equation}\label{eqn_of_state_parameter}
	\omega^{(\phi)}_{eff}=-1+x^2-\frac{\alpha  x z \left(\sqrt{1-x^2}-y^2\right)}{3 y^2 (1-z)},
\end{equation}
and the effective equation of state parameter for dark matter takes the form:
\begin{equation}
	\omega^{(m)}_{eff}=\frac{\alpha  x z}{3 (1-z)}.
\end{equation}
The global effective equation of state parameter for the model is expressed as:
\begin{equation}\label{eff_eqn_of_state_parameter}
	\omega_{eff}=-y^{2}\sqrt{1-x^2},
\end{equation}
and the deceleration parameter for the model is written as 
\begin{equation}\label{dec_parameter}
	q=\frac{1}{2} \left(1-3 y^{2}\sqrt{1-x^2} \right)
\end{equation}
From the above, one can achieve the condition for acceleration of the universe when:~$q<0$,~i.e.,~when $\omega_{eff}<-\frac{1}{3}$
and for deceleration, one follows the condition: $q>0$, i.e., $\omega_{eff}>-\frac{1}{3}.$
Friedmann eq. (\ref{Friedmann}) will give the constraint equation of $\Omega_{m}$ explicitly depending on the variables $x$ and $y$ as
\begin{equation}\label{density_dm}
	\Omega_{m}=1-\frac{y^{2}}{\sqrt{1-x^{2}}}
\end{equation}
which due to the energy condition $ 0\leq \Omega_{m} \leq 1 $ will lead to the following compact phase space:
\begin{equation}
	\Psi=\{ 0\leq x^{2}+y^4 \leq 1,~~ 0\leqslant z \leqslant1 \}
\end{equation} 


\section{Phase space analysis of autonomous system (\ref{autonomous_system 2}):}
\label{phase space autonomous system}
The critical points for the system (\ref{autonomous_system 2}) are the following
{\bf 
	\begin{itemize}
		\item  I. Set of critical points: $A=(x_{c},~0,~0)$
		\item  II. Critical point : $ B=(0,~1,~0)$
		
		\item  III. Critical point : $C=(0,~-1,~0)$
		
		\item  IV. Set of critical points: $ D=(1,~0,~z_{c})$
		
		\item  V. Set of critical points: $ E=(-1,~0,~z_{c})$
		
		\item  VI. Critical point : $ F=(0,~\sqrt{\frac{\alpha}{\alpha+\beta}},~1)$
		
		\item  VII. Critical point : $ G=(0,~-\sqrt{\frac{\alpha}{\alpha+\beta}},~1)$
	\end{itemize}
	
}
For autonomous system (\ref{autonomous_system 2}), Critical points and their corresponding physical parameters are shown in the table \ref{physical_parameters 1}.

	\begin{table}[tbp] \centering
		\caption{The Critical points and the corresponding physical parameters for the system (\ref{autonomous_system 2})  are presented }%
		\begin{tabular}
			[c]{cccccccc}\hline\hline
			\textbf{Critical Points}&$\mathbf{\Omega_{m}}$& $\mathbf{\Omega_{\phi}}$ & $\mathbf{\omega_{eff}^{(m)}}$ &
			$\mathbf{\omega^{(\phi)}_{eff}}$ & $\mathbf{\omega_{eff}}$ & $q$ &
			\\\hline
			$A  $ & $1$ & $0$ &
			$0$ & $\frac{0}{0}$ & $0$ & $\frac{1}{2}$ \\
			$B  $ & $0$ & $1$ &
			$0$ & $-1$ & $-1$ & $-1$\\
			$C $ & $0$ & $1$ &
			$0$ & $-1$ & $-1$ & $-1$ \\
			$D $ & $1$ & $0$ &
			$\frac{\alpha z_{c}}{3(1-z_{c})}$ & $\frac{0}{0}$ & $0$ & $\frac{1}{2}$\\
			$E  $ &  $1$ & $0$ &
			$-\frac{\alpha z_{c}}{3(1-z_{c})}$ & $\frac{0}{0}$ & $0$ & $\frac{1}{2}$\\
			$F  $ & $\frac{\beta}{\alpha+\beta}$ & $\frac{\alpha}{\alpha+\beta}$ & $\frac{0}{0}$ & $\frac{0}{0}$ & $-\frac{\alpha}{\alpha+\beta}$ & $\frac{\beta-2 \alpha}{2(\alpha+\beta)}$ \\
			$G  $ & $\frac{\beta}{\alpha+\beta}$ & $\frac{\alpha}{\alpha+\beta}$ & $\frac{0}{0}$ & $\frac{0}{0}$ & $-\frac{\alpha}{\alpha+\beta}$ & $\frac{\beta-2 \alpha}{2(\alpha+\beta)}$ \\
			
			\\\hline\hline
		\end{tabular}
		\label{physical_parameters 1} \\
		
\end{table}%
%
For the stability analysis, we have to  find out the corresponding eigenvalues of critical points and so, we present the eigenvalues for this model with
exponential potential  in  tabular form (see table \ref{eigenvalues1}).

	\begin{table}[tbp] \centering
		\caption{The eigenvalues of the linearized system (\ref{autonomous_system 2})}%
		\begin{tabular}
			[c]{ccccccc}\hline\hline
			\textbf{Critical Points} & $\mathbf{\lambda_{1}}$ &
			$\mathbf{\lambda_{2}}$ & $\mathbf{\lambda_{3}}$ \\\hline
			$ A $ & $ 0 $ & $0$ & $ 0 $  \\
			$ B $ & $ 0 $ & $-3$ & $-3$  \\
			$ C $ & $ 0 $ & $-3$ & $-3$  \\
			$ D $ & $ 0$ & $0$ & $0$  \\
			$ E $ &  $0$ & $0$ & $0$  \\
			$ F $ & $ 0 $ & $ -i \sqrt{\alpha } \sqrt{\beta } \sqrt{\frac{\alpha }{\alpha +\beta }}$ & $ i \sqrt{\alpha } \sqrt{\beta } \sqrt{\frac{\alpha }{\alpha +\beta }}$  \\
			$ G $ & $ 0 $ & $ -i \sqrt{\alpha } \sqrt{\beta } \sqrt{\frac{\alpha }{\alpha +\beta }} $ & $ i \sqrt{\alpha } \sqrt{\beta } \sqrt{\frac{\alpha }{\alpha +\beta }}$ 
			
			\\\hline\hline
		\end{tabular}
		\label{eigenvalues1}
\end{table}%

The table \ref{eigenvalues1} shows the eigenvalues  of the Jacobian matrix (evaluated at the critical points) for the  autonomous system (\ref{autonomous_system 2}) and we note that all the critical critical points are non-hyperbolic in nature.  So we can not analyze them by linear stability theory.  We need to use center manifold theory for some cases.  But center manifold theory can be used only for the critical points $B$ and $C$ as rest of the critical points do not have any eigenvalues with nonzero real part.  To analyze $A,D,E, F, G$ we use numerical procedure.\\

$\bullet$ The set of critical points $A$,$D$ and $E$ exist for all $\beta \in \mathbb{R}$ and $\alpha$.  $A$,$D$ and $E$ are lines of non isolated critical points.  These critical points describe the matter dominated era of
evolution. They are completely DM dominated solutions ($\Omega_m=1$), where DM corresponds to the form of dust ($\omega^{(m)}_{eff}=0$).  For this case, the DM can describe the decelerating phase of the evolution of the universe and coincidence problem cannot be alleviated by these points.  All the eigenvalues of the Jacobian matrix at $A$,$D$ and $E$ are zero.  So there is no analytical method to find the stability of these critical points.  We find the stability near $A$,$D$ and $E$ numerically and plot the vector field.  

\begin{figure}
	\centering
	\subfigure[]{
		\includegraphics[width=0.46\textwidth]{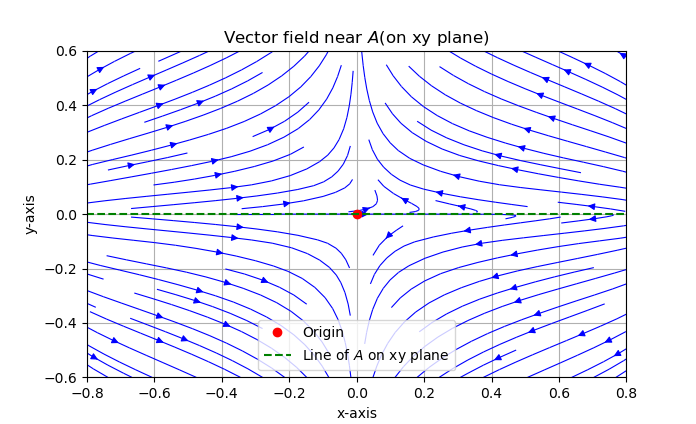}\label{Aa}}
	\subfigure[]{	
		\includegraphics[width=0.46\textwidth]{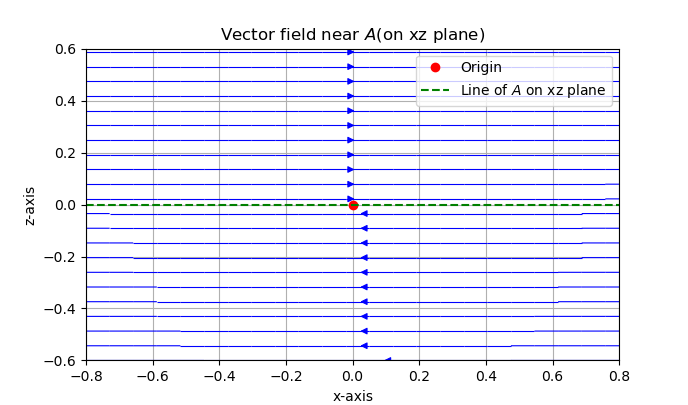}\label{Ab}}
	\caption{The figure shows stability of critical point $A$.  The green dotted line represents the non-isolated critical point $A$.  In panel (a) and (b),  the vector fields on the $xy$ plane and $xz$ plane are unstable near the line of critical points.}
	
	\label{A}
\end{figure}

\begin{figure}
	\centering
	\subfigure[]{
		\includegraphics[width=0.46\textwidth]{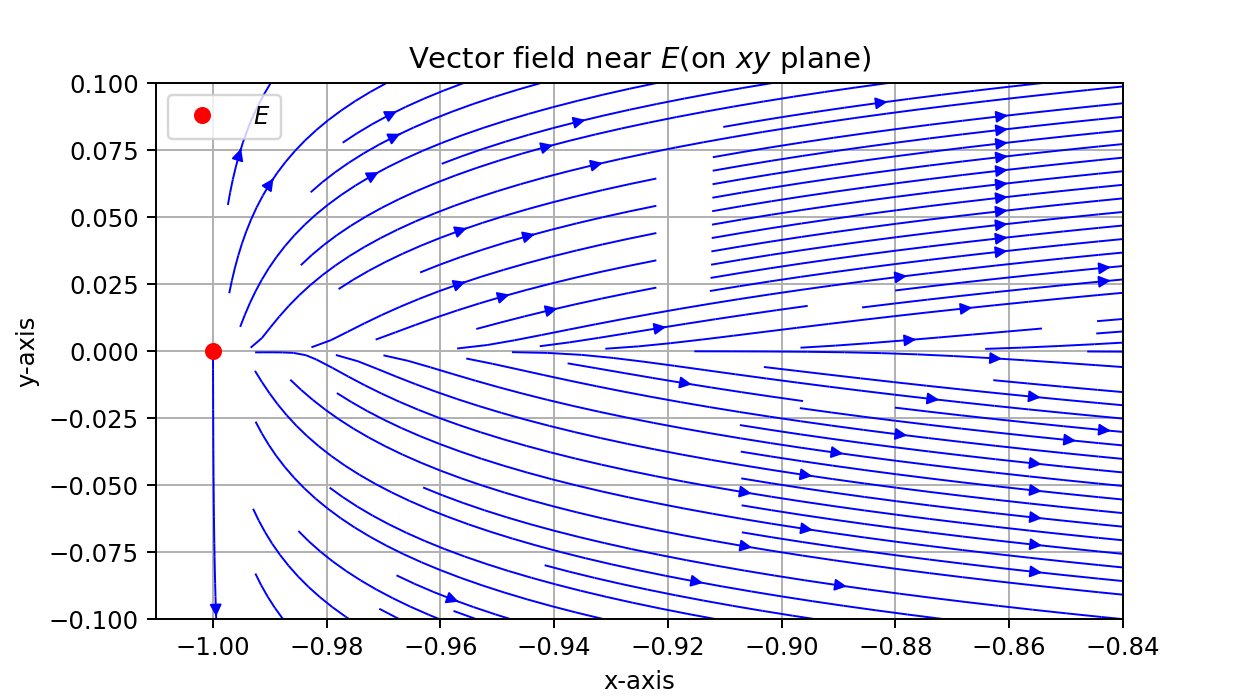}\label{Ea}}
	\subfigure[]{
		\includegraphics[width=0.46\textwidth]{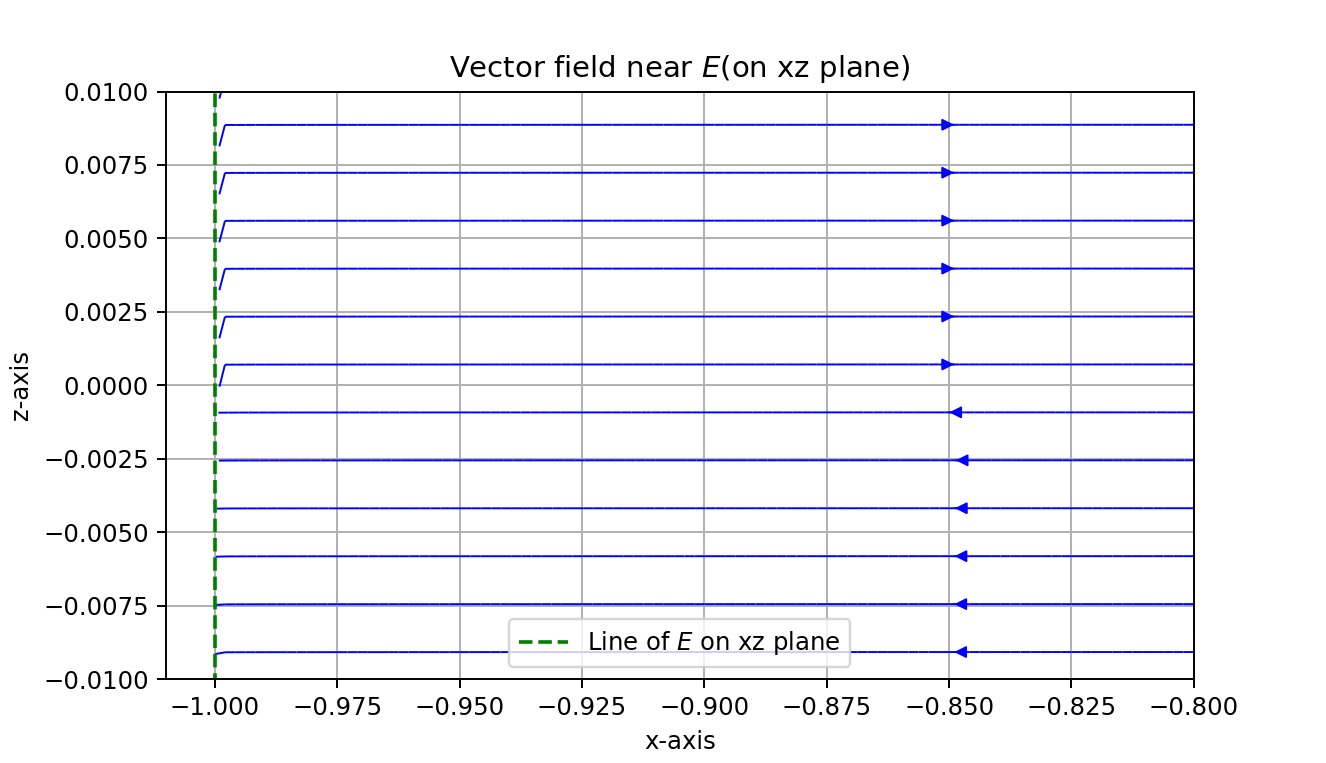}\label{Eb}}
	\caption{The figure shows stability of critical point $E$.  In panel (a),  the critical point on $xy$ plane is represented by red dot.  The vector field is repelling along $x$-axis and $y$-axis.  In panel (b),  the line $(-1,0,z_c)$ is represented by green dotted line.  The line $(-1,0,z_c)$ is an attractor for $z_c>0$ and repeller $z_c<0$ on the $xz$ plane. }
	
	\label{E}
\end{figure}

It is to be noted that the origin is a saddle on the $xy$ plane (see figure \ref{Aa}) and the vector field near $x$-axis ($A$) is unstable on both $xy$ and $xz$ plane (see figure \ref{A}).  The vector field near the  critical point $D$ on the $xy$-plane behaves as unstable node.  The flow is repelling along $x$-axis and $y$-axis on the $xy$ plane.  On the other hand,  the line $(1,0,z_c)$ is an attractor for $z_c>0$ and repeller for $z_c<0$ on the $xz$ plane.  But the line $(1,0,z_c)$ is a repeller on the $yz$ plane.   In figure (\ref{E}) it is to be noted that the vector field on the $xy$ plane near the critical point $E$ is unstable  in nature.  On the $xz$ plane the vector field is attracting for $z_c<0$ and repelling for $z_c>0$.

\begin{figure}
	\centering
	\subfigure[]{
		\includegraphics[width=0.3\textwidth]{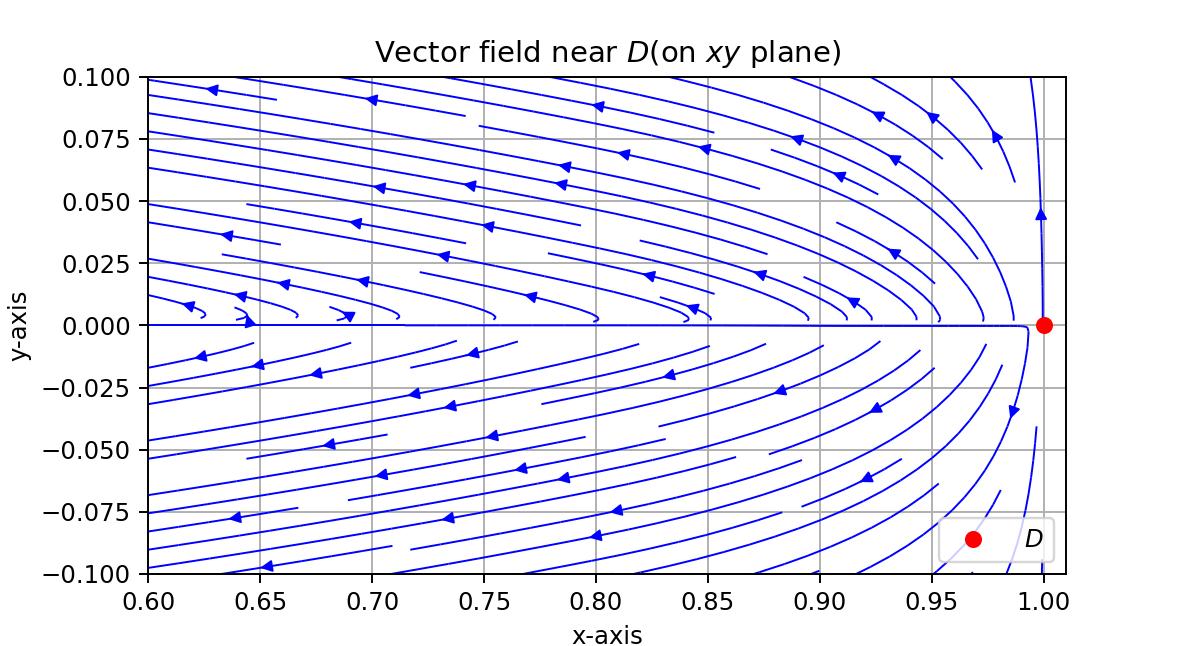}\label{Da}}
	\subfigure[]{
		\includegraphics[width=0.3\textwidth]{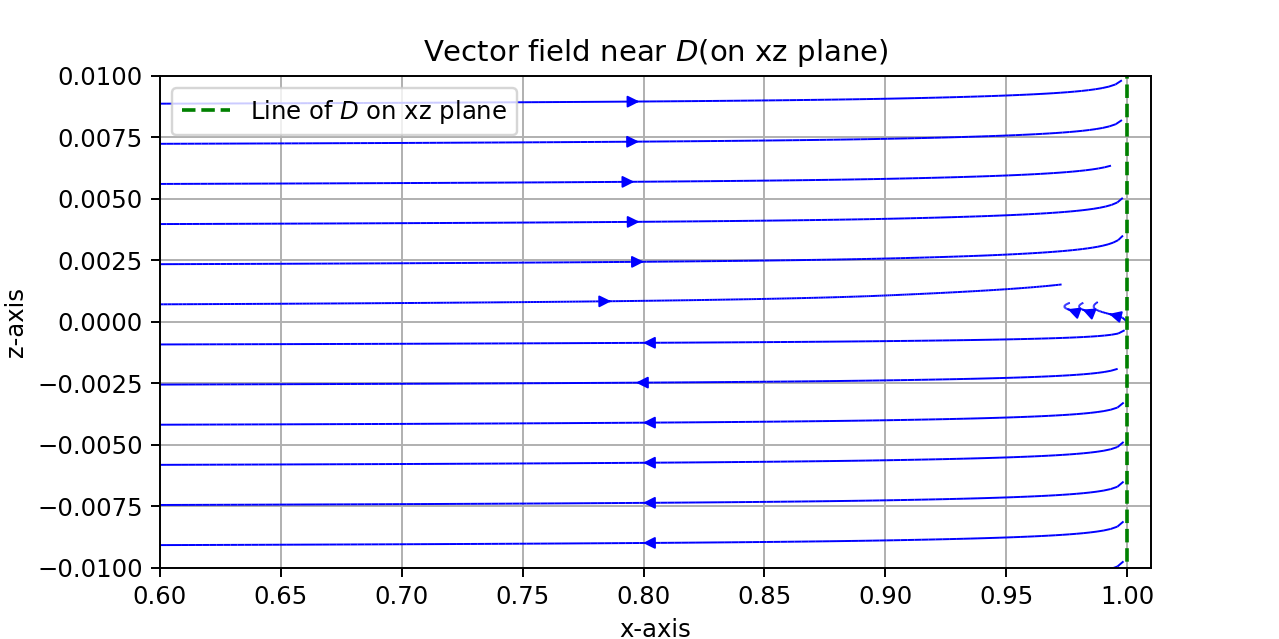}\label{Db}}
	\subfigure[]{
		\includegraphics[width=0.3\textwidth]{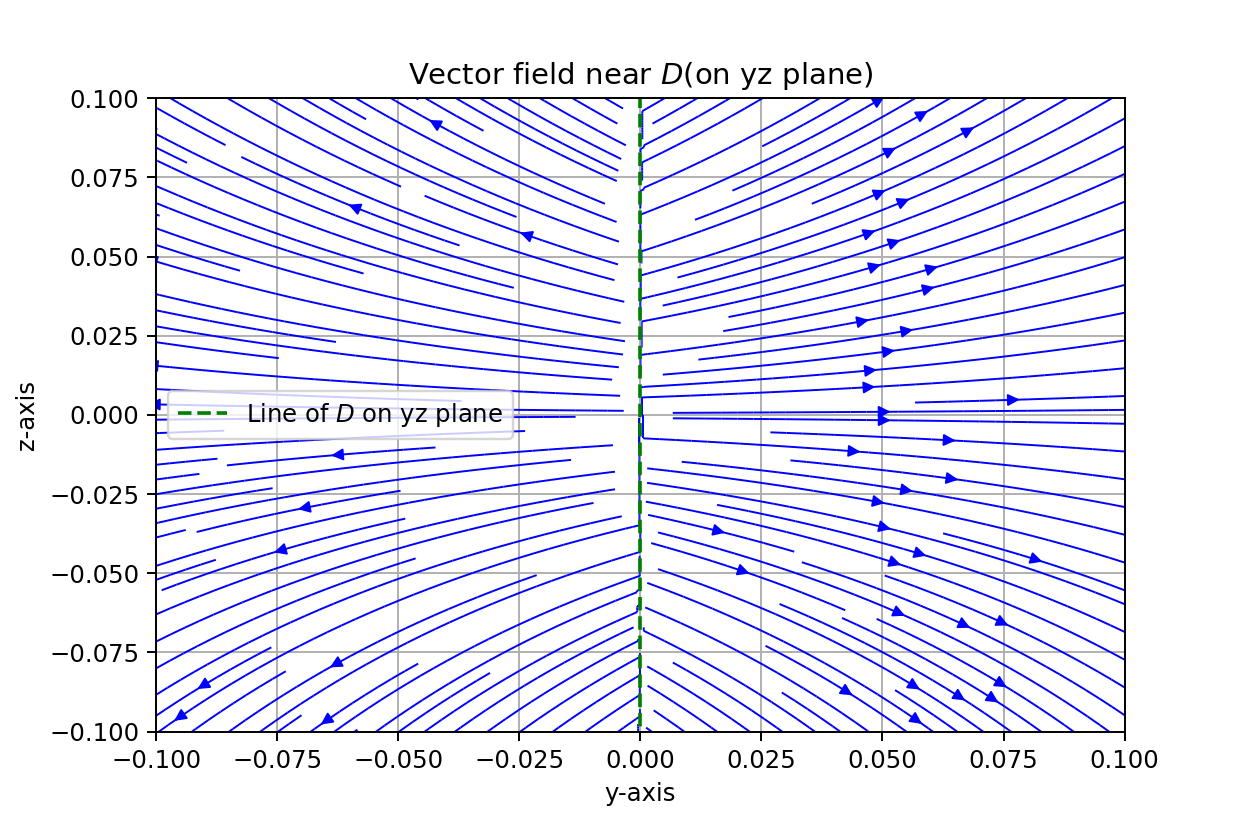}\label{Dc}}
	\caption{ The figure shows stability of critical point $D$.  In panel (a),  the critical point on $xy$ plane is represented by red dot.  The vector field is repelling along $x$-axis and $y$-axis.  In panel (b),  the line $(1,0,z_c)$ is represented by green dotted line.  The line $(1,0,z_c)$ is an attractor for $z_c<0$ and repeller $z_c>0$ on the $xz$ plane.   In panel (c),  the line $(1,0,z_c)$ is represented by green dotted line.   The line of critical points is a repeller on the $yz$ plane.}
	
	\label{D}
\end{figure}

$\bullet$ The set of critical points $B$ and $C$ exist for all $\beta \in \mathbb{R}$ and $\alpha$.   These
solutions are completely dominated by tachyon scalar field $(\Omega_{\phi}=1)$ which  behaves as cosmological constant ($\omega^{\phi}_{eff}=-1$).  The cosmic evolution near the points characterizes the de-Sitter expansion ($q=-1$) of the universe.  These critical points are isolated non-hyperbolic in nature and the Jacobian matrix at $B$ or $C$ contains only one vanishing eigenvalue.  We can employ center manifold theory (see for instance, \cite{Chakraborty2024,Mishra2019a,Boehmer2012,Perko} ) to find the characteristic of the vector field near these critical points as the following:

First we shift the point $B$ to the origin.  For that we take the change of variables as follows:\\
$$x \rightarrow x, ~y \rightarrow y+1,~z \rightarrow z. $$ 

The equation (\ref{autonomous_system 2}) reduces to 
\begin{subequations}
	\begin{align}
		\frac{dx}{dN}&=\left(1-x^2\right) \left(\alpha  \sqrt{1-x^2} z+3 x (y+1)^2 (z-1)-(y+1)^2 z (\alpha +\beta )\right).\label{auto_2Ba}\\
		\frac{dy}{dN}&=\frac{1}{2} (y+1)^3 \left(3 (1-z) \left(1-\sqrt{1-x^2} (y+1)^2\right)+\beta  x z\right).\label{auto_2Bb}\\
		\frac{dz}{dN}&=\frac{3}{2} (y+1)^2 (z-1)^2 z \left(1-\sqrt{1-x^2} (y+1)^2\right).\label{auto_2Bc}
	\end{align}
\end{subequations}
The Jacobian matrix of the system (\ref{auto_2Ba}-\ref{auto_2Bc}) can be obtained as
$$ J(x,y,z)=\begin{bmatrix}
	x z f-9 x^2 h (z-1)+3 h (z-1) & 2 \left(1-x^2\right) (y+1) (3 x (z-1)-z (\alpha +\beta )) & \left(1-x^2\right) \left(\alpha  g+3 x h-h (\alpha +\beta
	)\right) \\
	\frac{1}{2} (y+1)^3 \left(\beta  z-\frac{3 x h (z-1)}{g}\right) & \frac{3}{2} h \left(2gh (z-1)+3 (1-z) \left(1-gh\right)+\beta  x z\right) & \frac{1}{2} (y+1)^3 \left(\beta  x-3
	\left(1-gh\right)\right) \\
	\frac{3 x (y+1)^4 (z-1)^2 z}{2 g} & 3 (y+1) (z-1)^2 z \left(1-2 gh\right) & -\frac{3}{2} h (z-1) (3 z-1) \left(g y (y+2)+g-1\right) \\
\end{bmatrix}
$$
where, $f=\left(2 h (\alpha +\beta )-3 \alpha  g\right)$, $g=\sqrt{1-x^2}$, $h= (y+1)^2$
and at the origin the matrix takes the following simplified form
$$J(0,0,0)=\begin{bmatrix}
	-3 & 0 & -\beta\\
	0 & -3 & 0 \\
	0 & 0 & 0
\end{bmatrix}.$$
To find center manifold at the origin, we need to diagonalize the above Jacobian matrix.  For that we need to find $T$ such that $T^{-1}JT$ is a diagonal matrix.  So $T$ is given by
$$T=\begin{bmatrix}
	1 & 0 & -\frac{\beta}{3}\\
	0 & 1 & 0 \\
	0 & 0 & 1
\end{bmatrix}$$
If we consider J(0,0,0) as a linear transformation from $\mathbb{R}^3$ to $\mathbb{R}^3$ and the basis of $\mathbb{R}^3$ changes by $T$ matrix, then the coordinate vector (variables) $(x,y,z)$ transform to $(P,Q,R)$-coordinate system by the following rule
\begin{gather}
	\begin{bmatrix}
		x \\ y\\ z
	\end{bmatrix}=\begin{bmatrix}
		P-\frac{\beta}{3}R \\
		Q \\
		R
	\end{bmatrix}
\end{gather}
or equivalently,
\begin{gather}
	\begin{bmatrix}
		P \\ Q\\ R
	\end{bmatrix}=\begin{bmatrix}
		x+\frac{\beta}{3}z \\
		y \\
		z
	\end{bmatrix}
\end{gather}  
Now in the new $(P,Q,R)$-coordinate system the autonomous equations (\ref{auto_2Ba}-\ref{auto_2Bc}) take the following form:

\begin{multline}
	\frac{dP}{dN}=\frac{1}{2} \beta  (Q+1)^2 (R-1)^2 R \left(1-(Q+1)^2 \sqrt{1-\left(P-\frac{\beta  R}{3}\right)^2}\right)\\-\frac{1}{27} \left((\beta  R-3 P)^2-9 \right) \left(9 P (Q+1)^2 (R-1)
	+\alpha  R \sqrt{9-(\beta  R-3 P)^2}-3 (Q+1)^2 R (\alpha +\beta R)\right).\label{auto_3Ba} 
\end{multline}

\begin{align}
	\frac{dQ}{dN}&=\frac{1}{2} (Q+1)^3 \left(3 (1-R) \left(1-(Q+1)^2 \sqrt{1-\left(P-\frac{\beta  R}{3}\right)^2}\right)+\beta  R \left(P-\frac{\beta  R}{3}\right)\right).\label{auto_3Bb}\\
	\frac{dR}{dN}&=\frac{3}{2} (Q+1)^2 (R-1)^2 R \left(1-(Q+1)^2 \sqrt{1-\left(P-\frac{\beta  R}{3}\right)^2}\right).\label{auto_3Bc}  
\end{align}

Now near the origin higher degree terms of the equations of the vector field do not contribute significantly.  So ignoring the higher degree terms in the equations (\ref{auto_3Ba}-\ref{auto_3Bc}) one can obtain following simplified form
\begin{align}
	\frac{dP}{dN}&= -3P +3PR -6PQ -(2\alpha+\beta)QR  + \text{~higher degree terms}.\label{auto_4a}\\  
	\frac{dQ}{dN}&= -3Q -\frac{21}{2}Q^2 +3QR+\frac{\beta}{2}PR  +\text{~higher degree terms}.\label{auto_4b}\\
	\frac{dR}{dN}&= 6R^2Q -\frac{15}{2}Q^2R +\text{~higher degree terms}.\label{auto_4c}  
\end{align}

Then the flow along the center manifold  at $B$ is topologically equivalent (as $\det(T)>0$) to the flow of 
\begin{equation}
	\frac{dR}{dN} \approx 0. \label{CVF1}
\end{equation}

Similar to the above procedure, we shift the point $C$ to the origin.  For that we take the change of variables as follows:\\
$$x \rightarrow x, ~y \rightarrow y-1,~z \rightarrow z. $$ 
The equation (\ref{autonomous_system 2}) reduces to 
\begin{align}
	\frac{dx}{dN}&=\left(1-x^2\right) \left(\alpha  \sqrt{1-x^2} z+3 x (y-1)^2 (z-1)-(y-1)^2 z (\alpha +\beta )\right).\label{auto_2Ca}\\
	\frac{dy}{dN}&=\frac{1}{2} (y-1)^3 \left(3 (1-z) \left(1-\sqrt{1-x^2} (y-1)^2\right)+\beta  x z\right).\label{auto_2Cb}\\
	\frac{dz}{dN}&=\frac{3}{2} (y-1)^2 (1-z)^2 z \left(1-\sqrt{1-x^2} (y-1)^2\right).\label{auto_2Cc}
\end{align}
At the origin the jacobian matrix takes the following  form
$$J(0,0,0)=\begin{bmatrix}
	-3 & 0 & -\beta\\
	0 & -3 & 0 \\
	0 & 0 & 0
\end{bmatrix}.$$
Similar to the case of critical point $B$, we need the same $T$ to diagonalize $J(0,0,0)$ which is given by
$$T=\begin{bmatrix}
	1 & 0 & -\frac{\beta}{3}\\
	0 & 1 & 0 \\
	0 & 0 & 1
\end{bmatrix}$$
By the orientation preserving homeomorphism $T$ (as $\det(T)>0$), we get the new $(P,Q,R)$-coordinate system and the autonomous equations (\ref{auto_2Ca}-\ref{auto_2Cc}) take the following form:

\begin{multline}
	\frac{dP}{dN}=\frac{1}{2} \beta  (Q-1)^2 (R-1)^2 R \left(1-(Q-1)^2 \sqrt{1-\left(P-\frac{\beta  R}{3}\right)^2}\right)\\\frac{1}{27} \left((\beta  R-3 P)^2-9\right) \left(9 P (Q-1)^2 (R-1)+\alpha  R \sqrt{9-(\beta  R-3 P)^2}-3 (Q-1)^2 R
	(\alpha +\beta  R)\right).\label{auto_3Ca} 
\end{multline}

\begin{align}
	\frac{dQ}{dN}&=\frac{1}{2} (Q-1)^3 \left(3 (1-R) \left(1-(Q-1)^2 \sqrt{1-\left(P-\frac{\beta  R}{3}\right)^2}\right)+\beta  R
	\left(P-\frac{\beta  R}{3}\right)\right).\label{auto_3Cb}\\
	\frac{dR}{dN}&=\frac{3}{2} (Q-1)^2 (R-1)^2 R \left(1-(Q-1)^2 \sqrt{1-\left(P-\frac{\beta  R}{3}\right)^2}\right).\label{auto_3Cc}  
\end{align}
And ignoring the higher degree terms in the equations (\ref{auto_3Ca}-\ref{auto_3Cc}) one can obtain following simplified form
\begin{align}
	\frac{dP}{dN}&= -3P +3PR +6PQ +(2\alpha+\beta)QR  + \text{~higher degree terms}.\label{auto_4Ca}\\  
	\frac{dQ}{dN}&= -3Q +\frac{21}{2}Q^2 +3QR-\frac{\beta}{2}PR  +\text{~higher degree terms}.\label{auto_4Cb}\\
	\frac{dR}{dN}&= -6R^2Q -\frac{15}{2}Q^2R +\text{~higher degree terms}.\label{auto_4Cc}  
\end{align}

Then using center manifold theory we able to conclude that the flow along the center manifold  at $C$ is topologically equivalent to the flow of 
\begin{equation}
	\frac{dR}{dN} \approx 0. \label{CVF2}
\end{equation}

The vector fields of (\ref{CVF1}) and (\ref{CVF2}) are topologically conjugate near the critical points $B$ and $C$ respectively of the system (\ref{autonomous_system 2}).  The flow is almost static along the center manifold.  As a result of this, the vector field is dominated by the flow on $PR$-plane.  We also note that the nature of vector field does not depend on $\beta$ and the vector fields are stable near the critical points $B$ and $C$.  Thus the critical points  $B$ and $C$ are stable-node in nature and if the initial state is near $B$ or $C$ we have late time de-Sitter solution.

\begin{figure}
	\centering
	\subfigure[]{
		\includegraphics[width=0.46\textwidth]{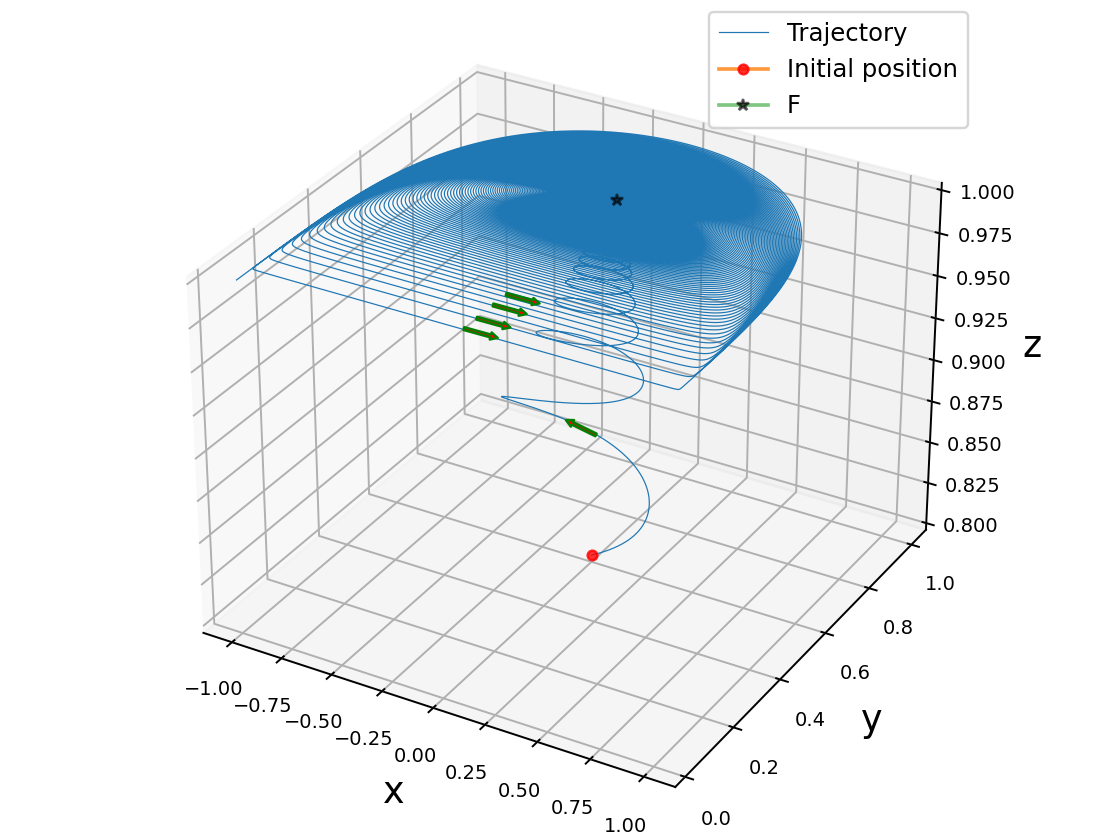}\label{fig:F}}
	\subfigure[]{
		\includegraphics[width=0.46\textwidth]{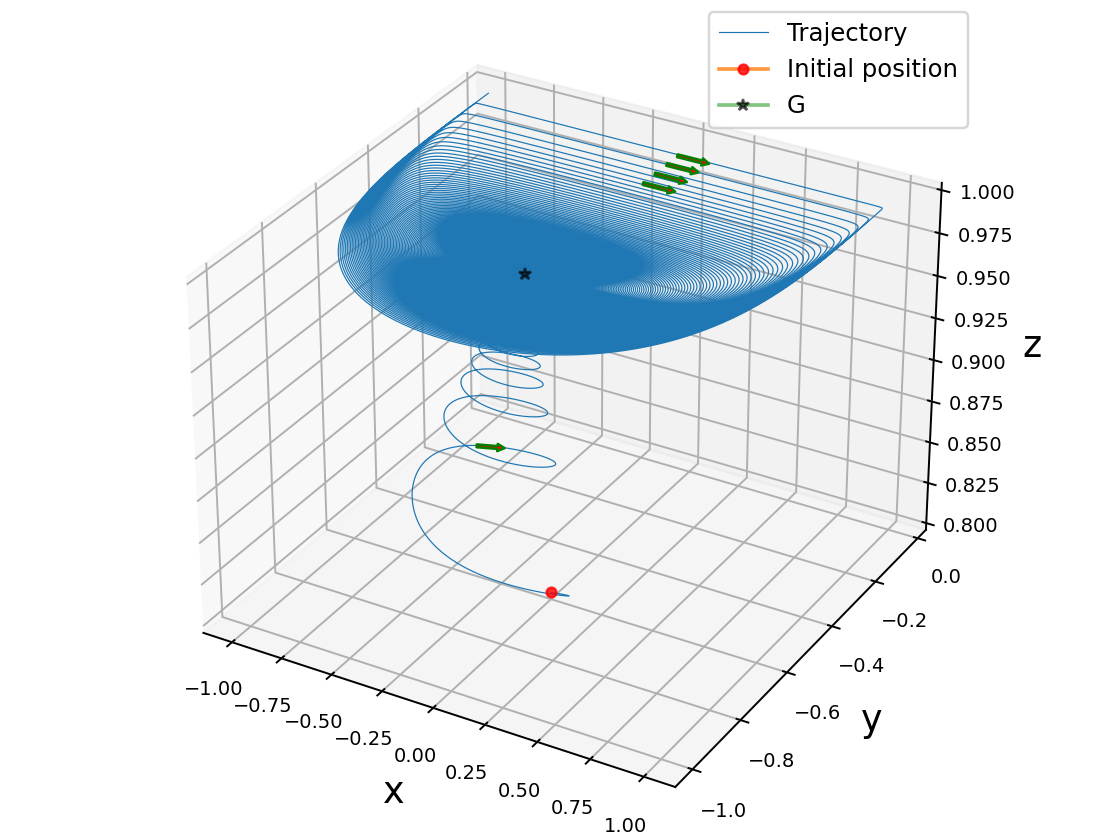}\label{fig:G}}
	\caption{The figure shows flow near the critical points $F$ and $G$ and it is to be noted that $z=1$ plane is an attractor near both the critical points $F$ and $G$ .  In panel (a),  trajectory starts from $(0.01,.6,0.9)$ near $F$ and goes towards $z=1$ plane.  The vector field flows anticlockwise along out going spiral around the critical point $F$ on $z=1$ plane. In panel (b), trajectory starts from $(0.01,-.6,0.9)$ near $G$ and goes towards $z=1$ plane.  The vector field flows clockwise around $G$.  In both cases the view is taken from the top of $z=1$ plane and we have taken $\alpha=\beta=1$. }
	\label{FG}
\end{figure}

$\bullet$ The critical points $F$ and $G$ exists for all $\alpha,~\beta >0$.  If we fluctuate $\alpha$ or $\beta$ we get non-isolated critical points $F$ and $G$.  These points correspond to the scaling solutions and the cosmic coincidence problem can be alleviated depending on the values of $\alpha$ and $\beta$.  Near these critical points dark matter and dark energy correspond to the form of dust and the evolution of the universe behaves as quintessence boundary for $2\alpha=\beta$.  On the other hand, there exists an accelerating universe when $2\alpha>\beta$.   The critical points $F$ and $G$ are situated on the $z=1$ plane and non-hyperbolic in nature.  As the real parts of all eigenvalues are zero, it is very difficult to find the character of the vector field around these points analytically.  We depict phase portrait numerically (see figures \ref{fig:F} and \ref{fig:G}) and try to understand the nature of flow near these points.   Now it is to be noted from our numerical analysis that $z=1$ plane is an attractor. The vector field is divided into two chambers by $xz$ plane.  If the initial position of the flow is in the positive (negative) $y$ coordinate but not on $z=1$ plane then the flow approaches towards $F$ (G).  But as soon as the flow touches the $z=1$ plane, it behaves as outgoing spiral.   The vector field flows anticlockwise along out going spiral around the critical point $F$ and  clockwise around $G$ when we see from the top of $z=1$ plane (see figures \ref{fig:F} and \ref{fig:G}). 

\begin{figure}
	\centering
	\subfigure[]{
		\includegraphics[width=0.44\textwidth]{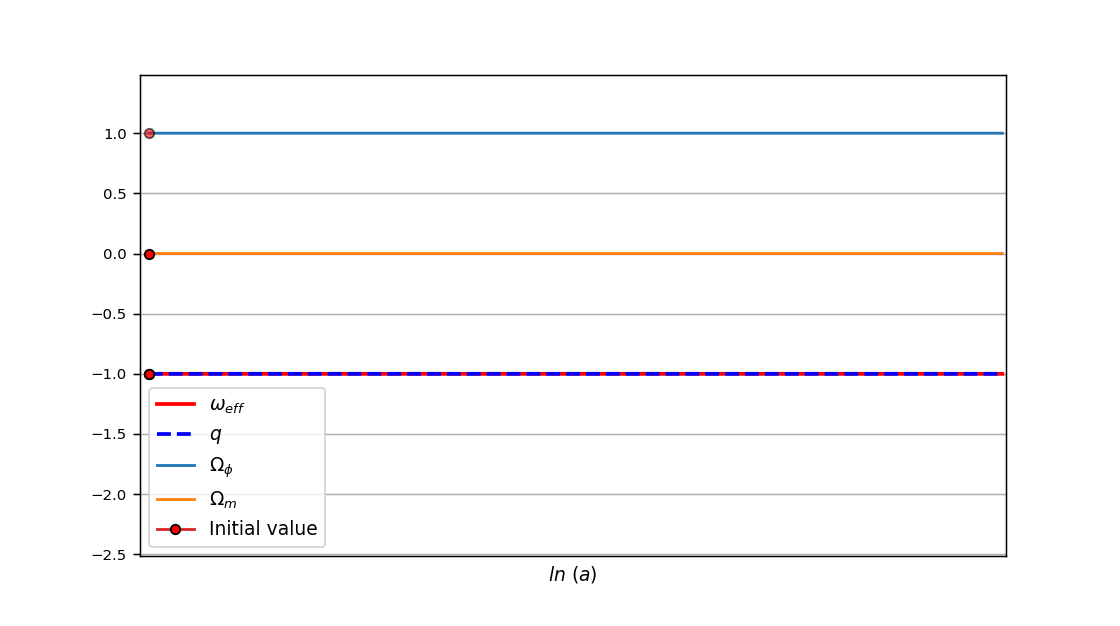}\label{BC_DP}}
	\subfigure[]{
		\includegraphics[width=0.44\textwidth]{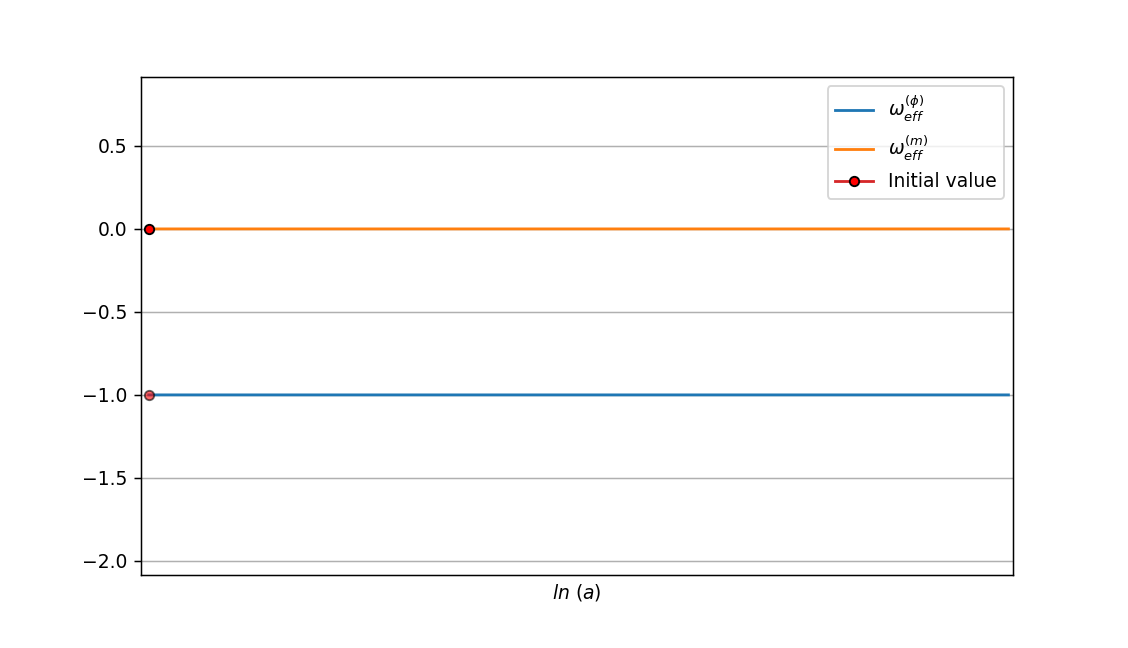}\label{BC_EDP}}
	\subfigure[]{
		\includegraphics[width=0.44\textwidth]{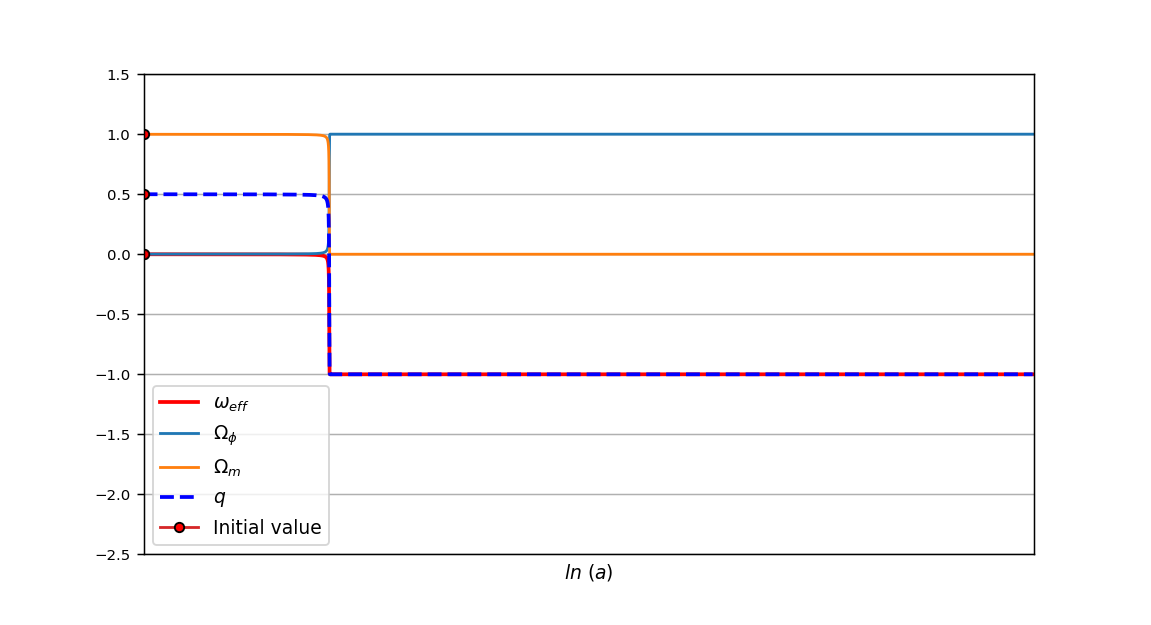}\label{DE_DP}}
	\subfigure[]{
		\includegraphics[width=0.44\textwidth]{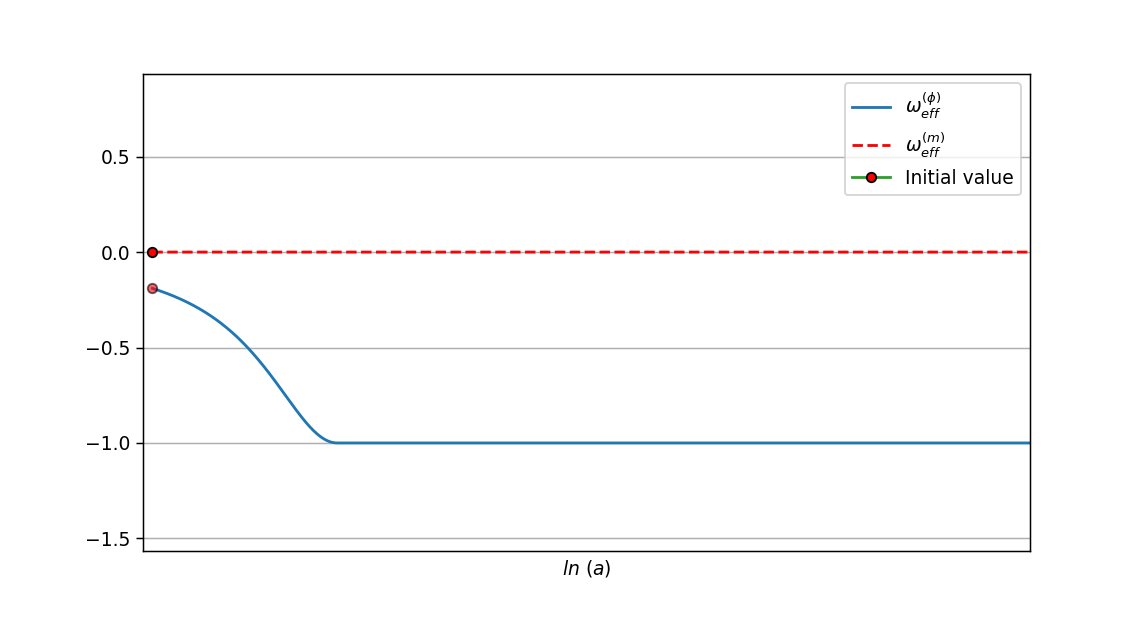}\label{DE_EDP}}
	\subfigure[]{
		\includegraphics[width=0.44\textwidth]{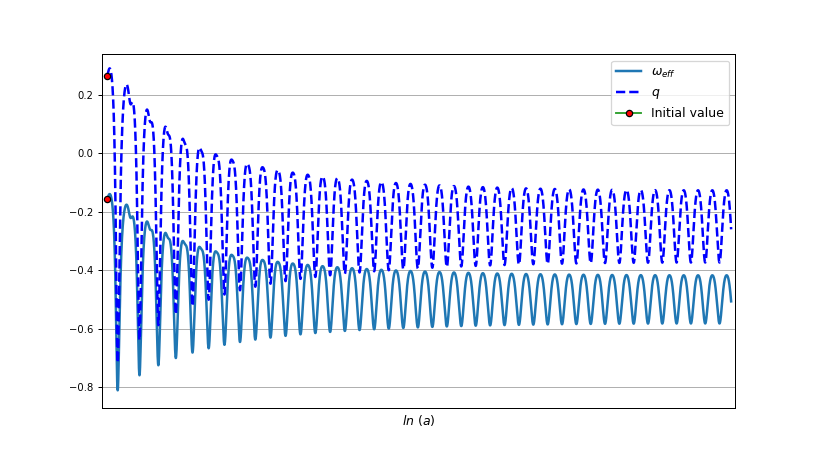}\label{FG_DP}}
	\subfigure[]{
		\includegraphics[width=0.44\textwidth]{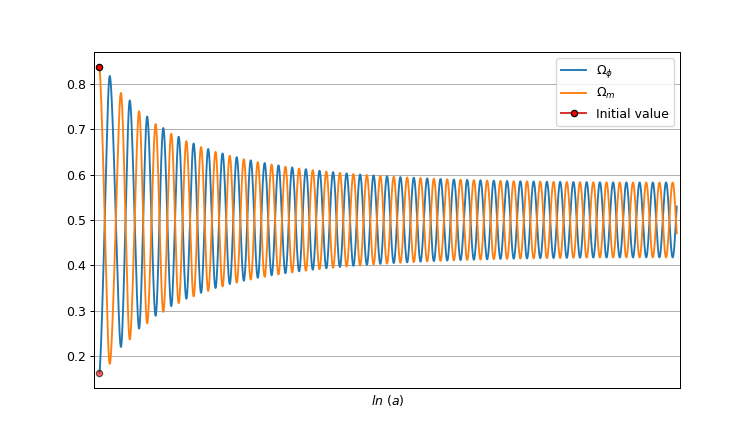}\label{FG_EDP}}
	\caption{The figures show the evolution of cosmological parameters numerically for different choices of initial conditions and $\alpha= \beta=1$. In panel (a) and (b), the starting position of the trajectory is near the critical point $B$ or $C$. In this case, we get scalar field dominated late-time solution. The cosmic evolution near the points characterizes the de-Sitter expansion of the universe. In panel (c) and (d), the starting position of the trajectory is near the critical point $D$ or $E$. There is a brief period of matter dominated deceleration expansion phase of the Universe before the transition to the scalar field dominated de-Sitter expansion of the universe. In panel (e) and (f), the starting position of the trajectory is near the critical point $F$ or $G$. As the trajectory goes towards $z=1$ plane in a spiral-like manner (see figure \ref{FG}), the values of cosmological parameters oscillate between a particular range. The nearer the trajectory goes towards $z=1$ plane the amplitude of oscillation  decreases and the values of deceleration parameter oscillate between $-0.18$ to $-0.38$ whereas the values of effective equation of state parameter oscillate between $-0.4$ to $-0.6$. On the other hand, the density parameters oscillate between $0.4$ to $0.6$ (a scaling solution) near the $z=1$ plane.}
	\label{Param}
\end{figure}

\section{Statefinder Diagnosis}\label{statefinder}

As mentioned in Introduction, different DE models have been proposed for interpreting the late-time observed cosmic acceleration in the literature. However, the problem of distinguishing among these models poses significant challenges. In this context, the authors of \cite{Sahni,Alam} have introduced a geometrical diagnostic pair ($r, s$), popularly known as a statefinder parameter. This parameter pair is geometric in nature as they rely on the scale factor of the universe directly. The ($r, s$) pair is defined as \cite{Sahni,Alam}
\begin{equation}
	r\equiv \frac{\dddot{a}}{a H^{3}},
\end{equation}
\begin{equation}
	s\equiv\frac{r-1}{3(q-\frac{1}{2})}\label{s}
\end{equation}
in which the over dot and $q =-\ddot{a}/aH^{2}$ denote the differentiation with respect to the cosmic time $t$ and  deceleration parameter, respectively. Different combinations of the pair ($r,s$) could serve as distinctive representation of different DE models.   If ($r=1,s=0$), then the DE behaves like cosmological constant (i.e., $\Lambda CDM$ model). On the other hand, ($r<1, s>0$) (or, $r=1, s=1$) suggests that the DE is a Quintessence (or, SCDM). For the case of Chaplygin gas model, the trajectories in the $s-r$ plane lie in the region where $s < 0$ and $r > 1$ \cite{Wu}. This makes it a powerful tool for effectively distinguishing between various DE models, even when they produce similar expansion histories. However, one can look into the refs. \cite{Sami,Myrzakulov,Rani,sfaaam,Setare,Zhang,Liu,Feng,Zhang1,Shao}, where a detailed statefinder pair analysis for different DE models is comprehensively discussed.\\

In this work, the present model is scrutinized employing statefinder diagnostic tools. The process involves calculating statefinder parameters with respect to redshift, compare them with that of $\Lambda CDM$, and exploring their behavior at both high and low redshift limits. Using the energy conservation eqs. (\ref{continuity DM}) and (\ref{continuity DE}), we obtain (after some calculations) the energy densities of DM and DE as 
\begin{equation}
	\rho_m\approx \rho_{m0}a^\eta \label{17}
\end{equation}
and
\begin{equation}
	\rho_\phi\approx a^{\eta-\mu}\rho_{\phi0}\left\{1-\frac{r_{0}\alpha \sqrt{\mu-\eta}}{\sqrt{3}\mu}a^{-\eta}\right\}\label{18}
\end{equation}
where $\eta = -3\left(\frac{\alpha}{3}\sqrt{1+\omega_{\phi}}+1 \right)$, $\mu = -3\left(\frac{\alpha}{3}\sqrt{1+\omega_{\phi}}-\omega_{\phi} \right)$, $r_0=\frac{\rho_{m0}}{\rho_{\phi0}}$ and $\omega_\phi=\frac{p_\phi}{\rho_\phi}$ is equation of state (EoS) parameter of DE. Also, $\rho_{m0}$ and $\rho_{\phi0}$ are considered as the present value of energy densities of DM and DE, respectively. Using the above expressions for energy densities ($\rho_{m}~\&~\rho_{\phi}$), the Friedmann equation in (\ref{Friedmann 1}) will give the analytic expression for the Hubble parameter as
\begin{equation}
	H(z)=(1+z)^{-\frac{\eta}{2}}H_0\left[\Omega_{m0}+\Omega_{\phi0}~(1+z)^\mu\left\{1-\frac{r_{0}\alpha \sqrt{\mu-\eta}}{\sqrt{3}\mu}(1+z)^{\eta}\right\}\right]^{\frac{1}{2}},\label{19}
\end{equation}
where $\Omega_{m0}+\Omega_{\phi 0}=1$ and $z=\frac{1}{a}-1$ is redshift parameter. For convenience, we introduce the dimensionless Hubble rate $E(z)\equiv H(z)/H_{0}$ and the parameters $\left\lbrace q,r,s\right\rbrace $, in terms of $E(z)$, can be written  as
\begin{equation}
	q(z)=-1+(1+z) \frac{E_{z}(z)}{E(z)},
\end{equation}
\begin{equation}
	r(z)=q(z)(1+2q(z))+(1+z)q_{z}(z).
\end{equation}
Here, the subscript $z$ stands for derivative with respect to $z$ and $s(z)$ is given by eq. (\ref{s}). For this model, we obtain

\begin{equation}
	\begin{split}
		E(z)=\frac{H}{H_{0}}=&(1+z)^{\frac{3}{2}\left(\frac{\alpha}{3}\sqrt{1+\omega_{\phi}}+1 \right)} \bigg[\Omega_{m0}+\Omega_{\phi0}~(1+z)^{-3\left(\frac{\alpha}{3}\sqrt{1+\omega_{\phi}}-\omega_{\phi} \right) } \\
		& \times \left\{1+\frac{\Omega_{m0}\alpha \sqrt{3(1+\omega_{\phi})}}{3\sqrt{3} \Omega_{\phi 0} \left(\frac{\alpha}{3}\sqrt{1+\omega_{\phi}}-\omega_{\phi} \right)}(1+z)^{ -3\left(\frac{\alpha}{3}\sqrt{1+\omega_{\phi}}+1 \right) }\right\} \bigg]^{\frac{1}{2}},
	\end{split}
\end{equation}

\begin{equation}
	q(z)=\frac{1}{2} \left(\alpha  \sqrt{\omega_{\phi}+1}+\frac{\Delta }{\Xi }+1\right),
\end{equation}
\begin{equation}
	\begin{split}
		r(z)=& \frac{1}{2} \bigg[\frac{\Omega_{m0} (z+1)^{-2 \alpha  \sqrt{\omega_{\phi}+1}+3 \omega_{\phi}-3} \left(\Omega_{\phi 0} (z+1)^{\alpha  \left(-\sqrt{\omega_{\phi}+1}\right)} \left(9 \omega_{\phi}^2 \Psi  \left(3 \omega_{\phi}-\alpha  \left(3 \sqrt{\omega_{\phi}+1}-\alpha \right)\right)+\Theta \right)+\chi \right)}{\Xi ^2 \left(3 \omega_{\phi}-\alpha  \sqrt{\omega_{\phi}+1}\right)} \\
		& +\left(\alpha  \sqrt{\omega_{\phi}+1}+\frac{\Delta }{\Xi }+1\right) \left(\alpha  \sqrt{\omega_{\phi}+1}+\frac{\Delta }{\Xi }+2\right)\bigg],
	\end{split}
\end{equation}
and
\begin{equation}
	\begin{split}
		s(z)=&\bigg[\frac{\Omega_{m0} (z+1)^{-2 \alpha  \sqrt{\omega_{\phi}+1}+3 \omega_{\phi}-3} \left(\Omega_{\phi0} (z+1)^{\alpha  \left(-\sqrt{\omega_{\phi}+1}\right)} \left(9 \omega_{\phi}^2 \Psi  \left(3 \omega_{\phi}-\alpha  \left(3 \sqrt{\omega_{\phi}+1}-\alpha \right)\right)+\Theta \right)+\chi \right)}{\Xi ^2 \left(3 \omega_{\phi}-\alpha  \sqrt{\omega_{\phi}+1}\right)} \\ & +\left(\alpha  \sqrt{\omega_{\phi}+1}+\frac{\Delta }{\Xi }+1\right) \left(\alpha  \sqrt{\omega_{\phi}+1}+\frac{\Delta }{\Xi }+2\right)-2 \bigg] \div {3 \left(\alpha  \sqrt{\omega_{\phi}+1}+\frac{\Delta }{\Xi }\right)}
	\end{split}
\end{equation}
where,
\begin{align*}
	\Delta =&(z+1)^{-2 \alpha  \sqrt{\omega_{\phi}+1}+3 \omega_{\phi}-3} \bigg(-\Omega_{m0} \alpha  \sqrt{\omega_{\phi}+1}+\frac{\Omega_{m0} \alpha  \left(\alpha +\alpha  \omega_{\phi}+3 \sqrt{\omega_{\phi}+1}\right)}{3 \omega_{\phi}-\alpha  \sqrt{\omega_{\phi}+1}}+\Omega_{\phi 0} \left(3 \omega_{\phi}-\alpha  \sqrt{\omega_{\phi}+1}\right) \\ 
	& \times (z+1)^{\alpha  \sqrt{\omega_{\phi}+1}+3}\bigg), \\
	\Xi =&\frac{\Omega_{m0} \alpha  \sqrt{\omega_{\phi}+1} (z+1)^{-2 \alpha  \sqrt{\omega_{\phi}+1}+3 \omega_{\phi}-3}}{\alpha  \sqrt{\omega_{\phi}+1}-3 \omega_{\phi}}+\Omega_{m0}+\Omega_{\phi0} (z+1)^{3 \omega_{\phi}-\alpha  \sqrt{\omega_{\phi}+1}}, \\
	\Psi =&(z+1)^{2 \alpha  \sqrt{\omega_{\phi}+1}+3}, \\
	\chi =&-\Omega_{m0} \alpha  \left(3 \omega_{\phi}^2 \left(3 \sqrt{\omega_{\phi}+1}-4 \alpha \right)+2 \left(2 \alpha ^2-9\right) \sqrt{\omega_{\phi}+1} \omega_{\phi}+4 \alpha  \left(\alpha  \sqrt{\omega_{\phi}+1}+3\right)+9 \sqrt{\omega_{\phi}+1}\right),  \mbox{and}~\\
	\Theta =&\alpha ^2 \omega_{\phi} \left(-(1 + z)^{3\omega_{\phi}}  \left(\alpha  \sqrt{\omega_{\phi}+1}+6\right)+\alpha  \left(-\sqrt{\omega_{\phi}+1}\right) \left(z^3+3 z^2+3 z+1\right) (z+1)^{2 \alpha  \sqrt{\omega_{\phi}+1}}+9 \Psi \right) \\
	&-\alpha  \left(\alpha  \left(\alpha  \sqrt{\omega_{\phi}+1} ((1 + z)^{3\omega_{\phi}} +\Psi )+6 (1 + z)^{3\omega_{\phi}} \right)+9 \sqrt{\omega_{\phi}+1} (1 + z)^{3\omega_{\phi}} \right). 
\end{align*}

\begin{figure}
	\centering
	\subfigure[]{%
		\includegraphics[width=7.2cm,height=6.3cm]{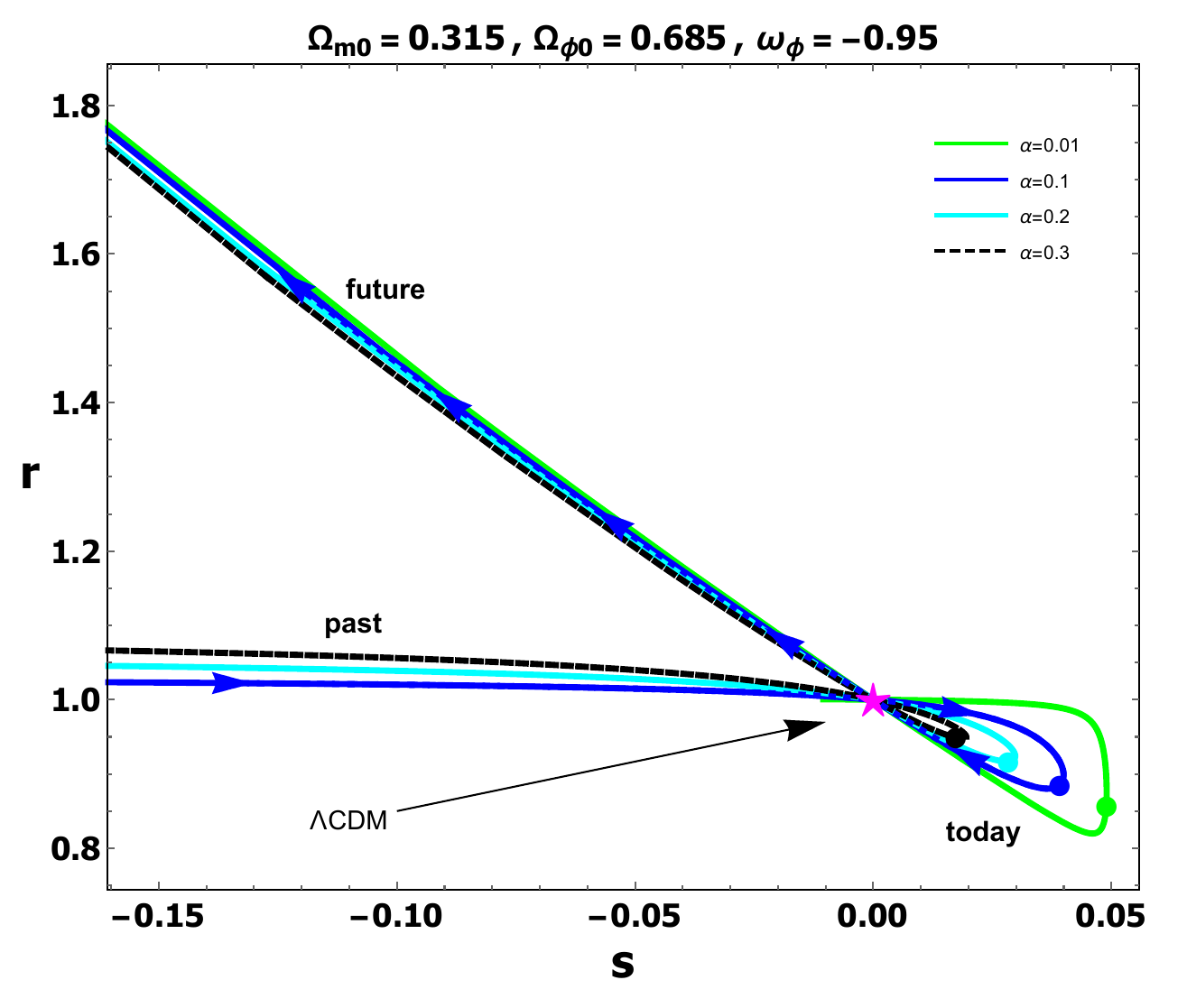}\label{sr-quintessence-alpha}}
	\qquad
	\subfigure[]{%
		\includegraphics[width=7.2cm,height=6.3cm]{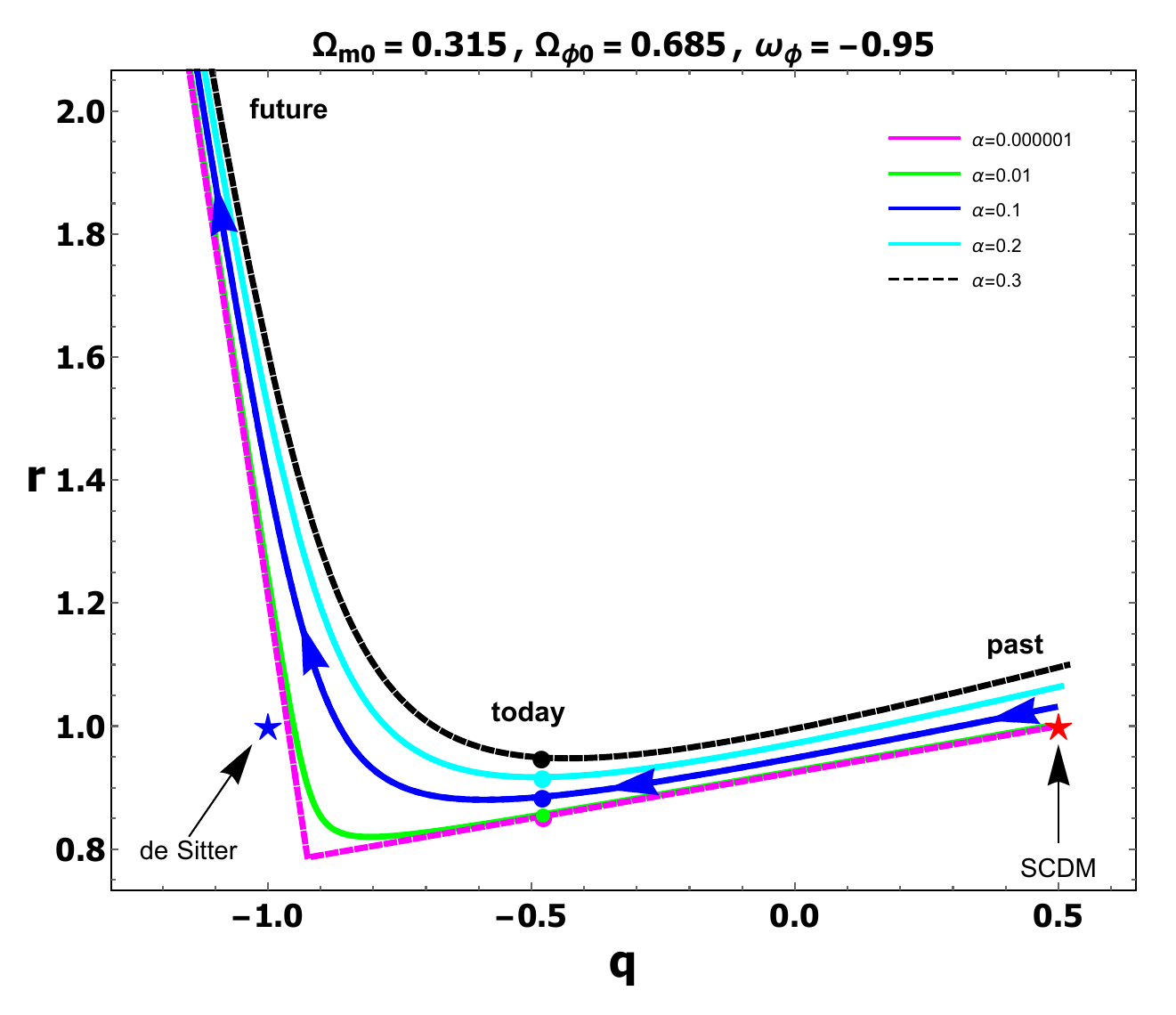}\label{qr-quintessence-alpha}}
	
	\caption{The time evolutions of the statefinder pairs in the $s-r$ plane (left panel) and $q-r$ plane (right panel) for this model are shown using different values of $\alpha$, as indicated in panel. The magenta colored star $(s = 0,r = 1)$ in the left panel corresponds to the $\Lambda CDM$ model, while in the right panel, the red colored star $(q = 0.5,r = 1)$ represents the matter dominated Universe $(SCDM)$ and the blue colored star $(r=1,q=-1)$ denotes the de Sitter fixed point. In each panel, the colored dots on the curves show present values of the pairs $(s_{0},r_{0})$ and $(q_{0},r_{0})$ for different values of $\alpha$.}
	\label{quintessence-alpha}
\end{figure}
\begin{figure}
	\centering
	\subfigure[]{%
		\includegraphics[width=7.2cm,height=6.3cm]{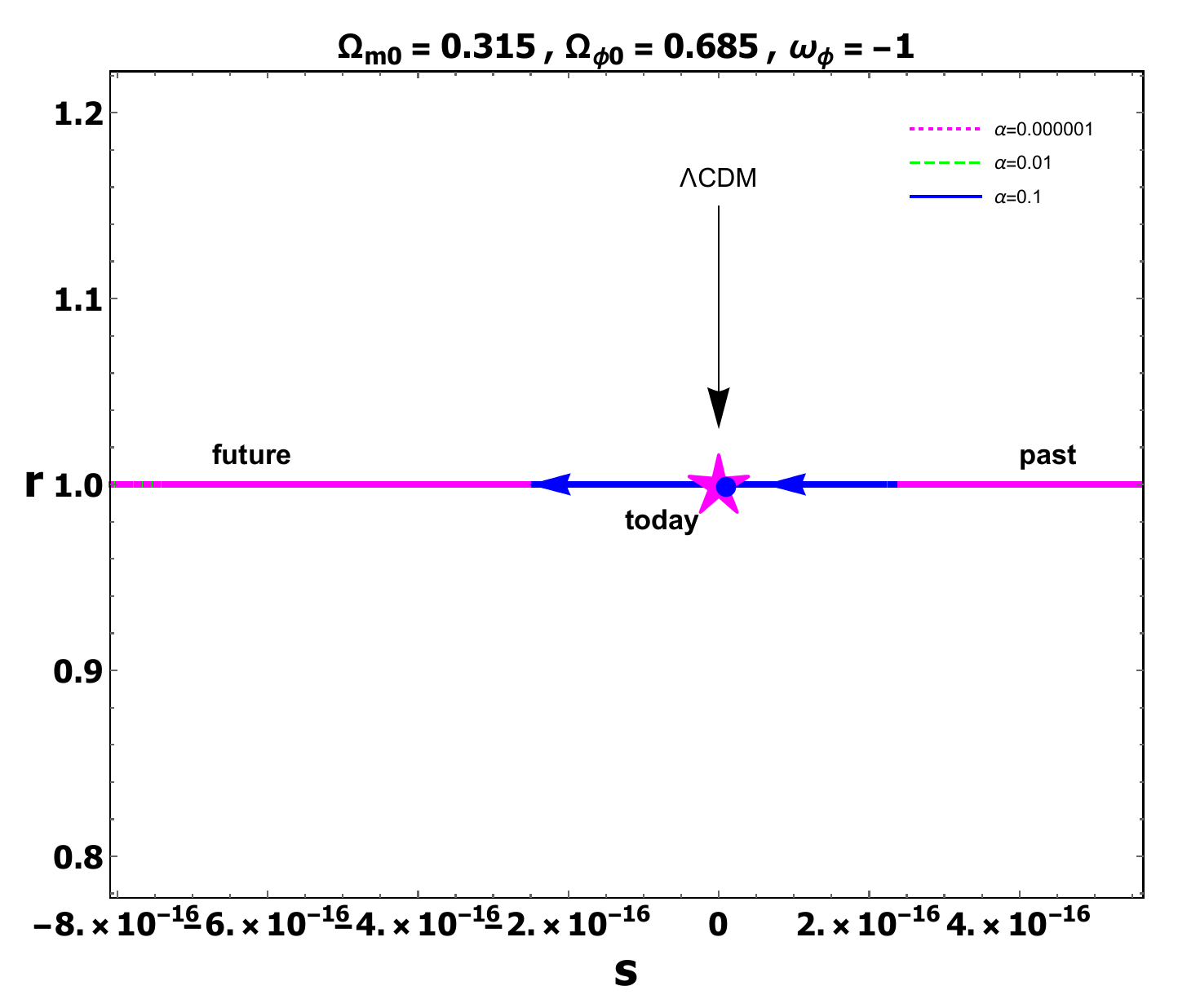}\label{sr-cc-alpha}}
	\qquad
	\subfigure[]{%
		\includegraphics[width=7.2cm,height=6.3cm]{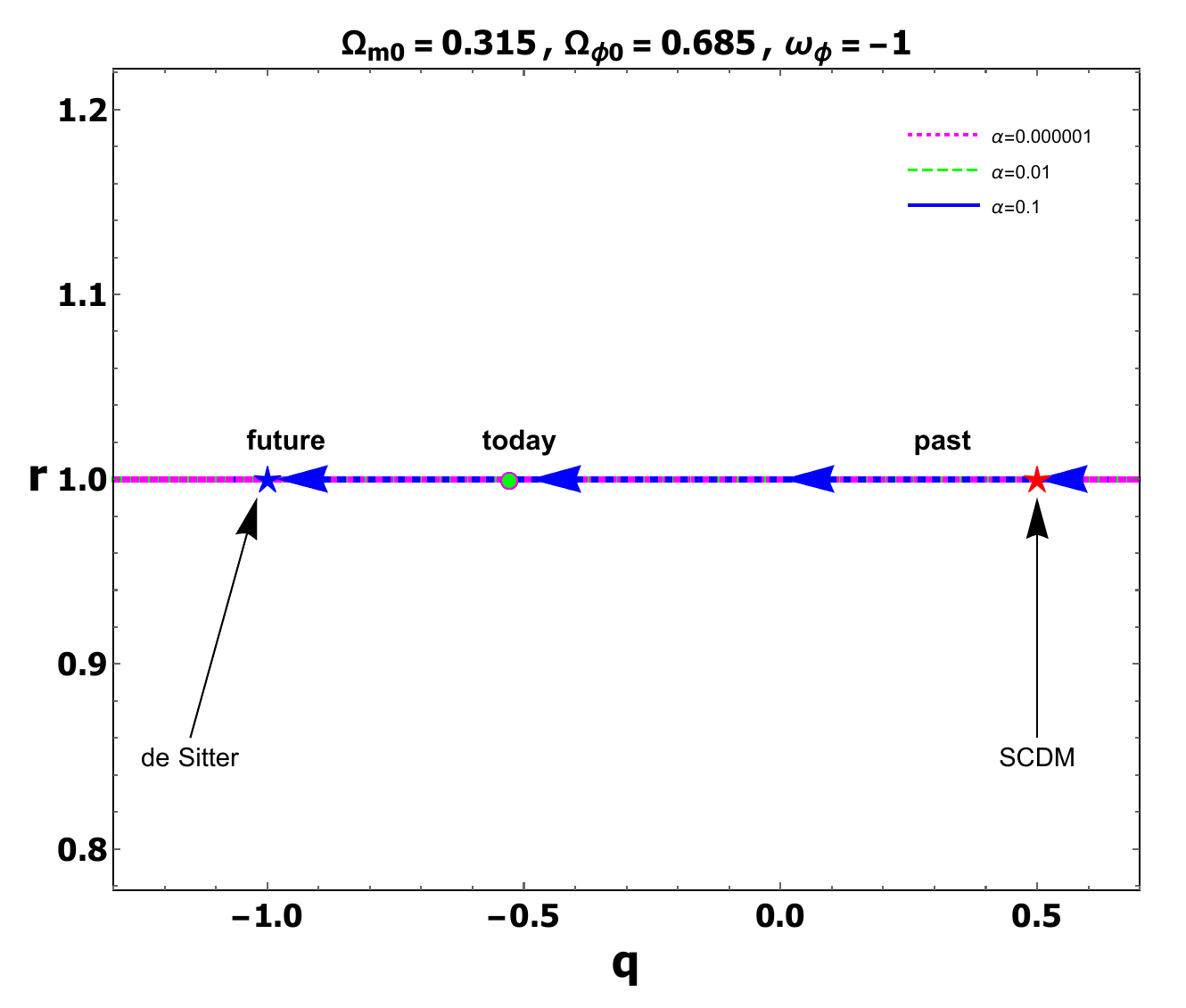}\label{qr-cc-alpha}}
	
	\caption{The time evolutions of the statefinder pair in the $s-r$ plane (left panel) and in the $q-r$ plane (right panel) for this model are shown. The magenta colored star $(s = 0,r = 1)$ in the left panel corresponds to the $\Lambda CDM$ model, while, in the right panel, the red colored star $(q = 0.5,r = 1)$ represents the matter dominated Universe $(SCDM)$ and the blue colored star $(r=1,q=-1)$ denotes the de Sitter fixed point. In each panel, the colored dots on the curves show present values of the pairs $(s_{0},r_{0})$ and $(q_{0},r_{0})$ for different values of $\alpha$.}
	\label{cc-alpha}
\end{figure}

The evolutionary trajectories of the statefinder pair in the $s$-$r$ plane are shown in figures \ref{sr-quintessence-alpha} for different choices of the model parameter $\alpha$ and $\omega_{\phi}=-0.95$. The evolutionary trajectories
of statefinder pair $(s,r)$ of our model start its evolution along the direction of $s$ increasing and $r$ decreasing and with time, trajectories pass through the $\Lambda CDM$ fixed point $(s=0,r=1)$. After making a swirl, these lies in the region $(s < 0,r > 1)$ in the future. Also for different values of the model parameter $\alpha$, we found different trajectories and present values of the statefinder pair $(s_{0},r_{0})$ as shown by the colored dots. This scenario shows that $\alpha$ affects on the evolutionary trajectories in the $s$-$r$ plane. Moreover, in figure \ref{sr-cc-alpha}, we obtain same evolutionary curve in $s$-$r$ plane for different values of the model parameter $\alpha$ and $\omega_{\phi}=-1$. In contrary to figure \ref{sr-quintessence-alpha}, we found that $\alpha$ does not affect on the evolutionary trajectories in the $s$-$r$ plane for the case $ \omega_\phi=-1$. Furthermore, we have also shown the evolutionary trajectories of another statefinder pair $(q,r)$ for the model in figures \ref{qr-quintessence-alpha} and \ref{qr-cc-alpha} . From figure \ref{qr-quintessence-alpha}, we found that for $\alpha=0.000001$, the evolutionary curve of statefinder pair $(q,r)$ of the model starts from the $SCDM$ $(r=1,q=0.5)$ in the past while the other evolution curves for the model have a smaller deviation from this fixed point and finally all the curves reach above the de Sitter expansion (SS) $(r=1,q=-1)$ in the future.  We also found from figure \ref{qr-cc-alpha} that $\alpha$ has no effect on the evolutionary trajectories in the $q$-$r$ plane for $\omega_\phi=-1$. Moreover, figures \ref{quintessence-alpha}, \ref{cc-alpha} clearly indicate that the evolution of $q$ shows a smooth signature flip from its positive value regime to negative value regime in the $q$-$r$ plane. Hence, from the statefinder diagnostics analysis, the present model explains the late-time accelerated universe and also the transition from the early decelerated phase ($q>0$) to the current accelerated phase ($q<0$).\\

\section{Comparison of the model with the observational data}\label{data}
\begin{figure}
	\centering
	\subfigure[]{%
		\includegraphics[width=7.2cm,height=6.3cm]{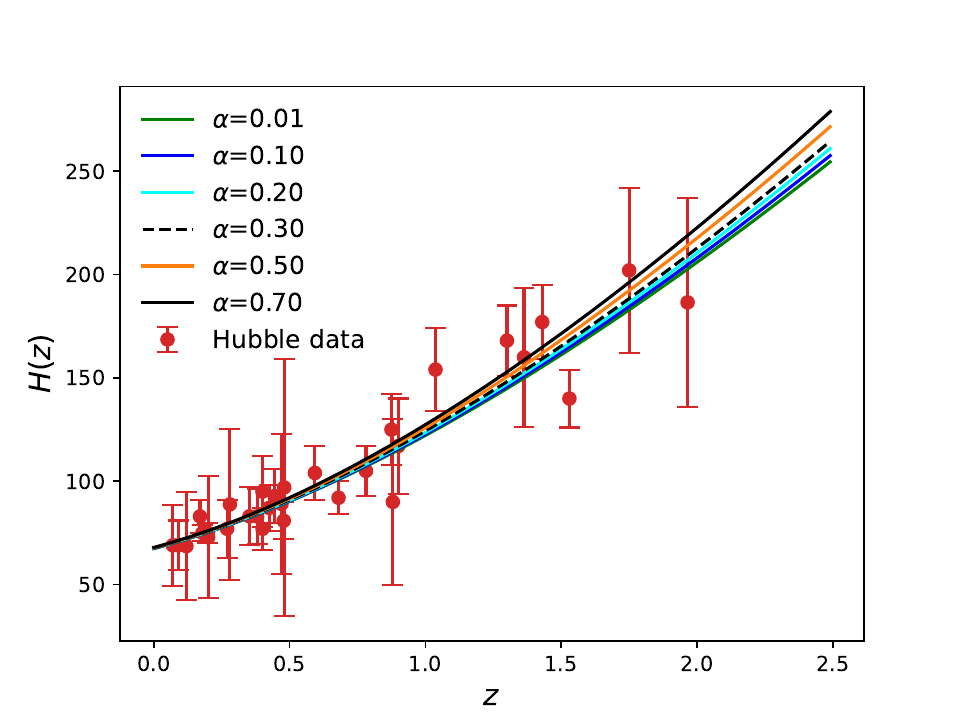}\label{data-h}}
	\qquad
	\subfigure[]{%
		\includegraphics[width=7.2cm,height=6.3cm]{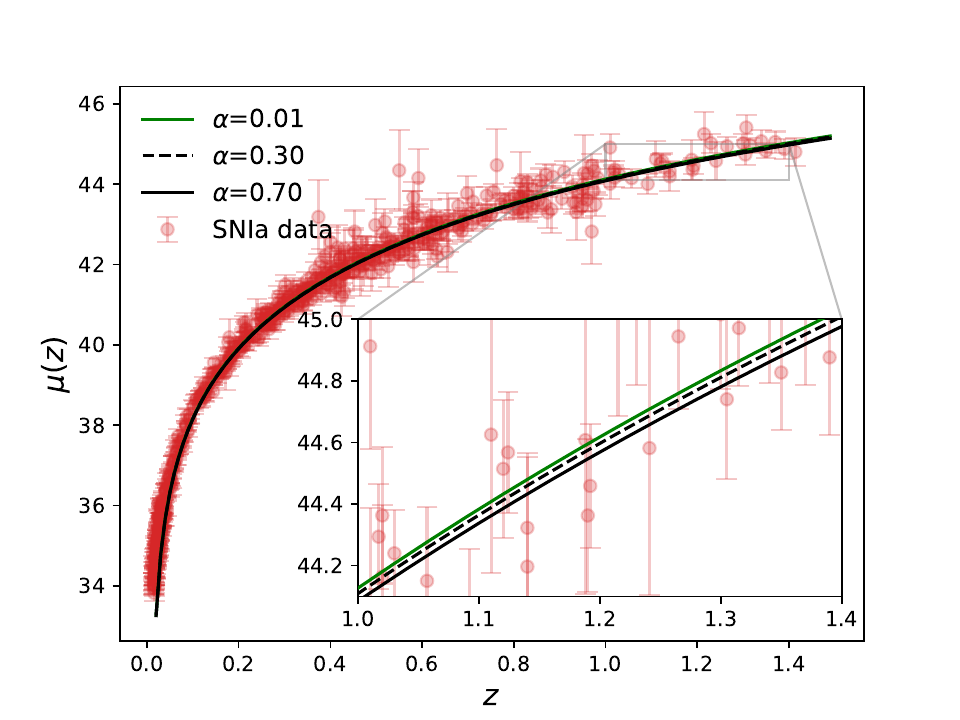}\label{data-sn}}
	
	\caption{In figure (a) - (b), the evolutions of the Hubble
		parameter $H(z)$ and distance modulus $\mu(z)$ are shown
		with respect to the redshift $z$ for the present model, and compared with 31 data points for Hubble parameter \cite{dH31} and 580 data points for Type Ia Supernova  \cite{dsnia} observations, respectively. This is for $\omega_{\phi}=-0.95$, $\Omega_{m0}=0.315$,  $H_{0}=67.4$ km/s/Mpc \cite{dH0} and different values of $\alpha$, as indicated in panel.}
	\label{datahsn}
\end{figure}

To check the reliability of our model further, the predicted evolution of the Hubble parameter and distance modulus have also been studied and compared that with the observational Hubble parameter and Type Ia Supernova (SNIa) datasets in figure \ref{datahsn}. In figure \ref{data-h}, we have plotted the nature of the Hubble parameter as a function of redshift $z$ and compared it with that of the Hubble data with $1\sigma$ error bars obtained from
the compilation of 31 points of $H(z)$ measurements \cite{dH31}. Also, the theoretical evolution of distance modulus, which is the difference between the apparent and
absolute magnitudes of the observed SNIa, as a function of $z$ using this model
is compared with the SNIa data points \cite{dsnia} and is shown in figure \ref{data-sn}. It is clear from figures \ref{data-h} and \ref{data-sn} that the theoretical model show proximity with observational results at low redshift very well.
%
%
%




%
%

\section{Concluding Remarks}\label{conclusion}
In this paper, we have studied a cosmological model of interacting tachyon with varying-mass dark matter particles in the background of FLRW universe where the tachyonic fluid plays the role of DE and pressureless dust as dark matter. In this variable mass particle (VAMP) mechanism, it is assumed that the mass of the DM particles is time dependent through the scalar field $\phi$ in the sense that the decaying of DM particles reproduces the scalar field. We have studied the cosmological model in the context of dynamical analysis by transforming suitable dimensionless variables. We considered the mass of DM particles and the scalar field potential to be varied with exponential function of $\phi$.  In this study we considered the exponential mass dependence and exponential potential function of scalar field $\phi$, i.e.,
$Q_{m}(\phi)=Q_{m0}~ \mbox{exp} \{-\alpha H_{0}\phi\}$ and $V(\phi)=V_{0}~exp\{\beta H_{0} \phi\}$, where $Q_{m0}$ and $V_{0}$ are constant and $\alpha$ and $\beta$ are constant parameters. As a result, the autonomous system is reduced to be a 3D system. We have studied the phase space analysis of the system (\ref{autonomous_system 2}). We have extracted seven critical points from this system. The critical points and the cosmological parameters are presented in the table \ref{physical_parameters 1} and the eigenvalues of the Jacobian matrix are displayed in the table \ref{eigenvalues1}. We observed that all the critical points are non-hyperbolic in nature. Among them, the set of critical points $A$, $D$ and $E$ are non-isolated sets having all the eigenvalues zero for which it is very hard to find the nature of these sets. However, numerical investigations are carried out to find the nature of these sets. These sets behave as dust dominated decelerated intermediate phase of the universe (since $\Omega_{m}=1$,~ $\omega_{eff}=0,~q=\frac{1}{2}$). These solutions are physically interested because they can successfully describe the phase of recent past evolution of the universe. On the other hand, the DE dominated solutions namely, the points $B$ and $C$ are non-hyperbolic having the values of cosmological parameters $\Omega_{\phi}=1,~\omega_{eff}=-1,~q=-1$ can show the de Sitter expansion of the universe. Due to physical importance, we have performed the center manifold theorem to study their stability. We note that the character of the flow on the center manifold does not depend on the parameter $\beta$ (see eqs. (\ref{CVF1}) and (\ref{CVF2})).  For the case of critical  points $B$ and $C$, the flow on the center manifold is topologically equivalent to $\frac{dZ}{dN} \approx \frac{3}{2}Z^3$.  It is to be noted that the vector field on the center manifold is repelling away from the critical points.  On the other hand, the vector field along the eigenvector of the eigenvalue $\lambda_{2}=-3$ is attracting towards the critical points.   So the late-time solutions near the critical points $B$ and $C$ depend on initial conditions \cite{Mishra2019}. Finally, we obtained the DE-DM scaling solutions described by the critical points $F$ and $G$ which are also non-hyperbolic points in the phase space and due to complicated nature we have performed numerical investigation in figure \ref{FG} to show their stability. We can conclude that the points can show the late time scaling attractors solving the coincidence problem. If the trajectory of the vector field in the phase-space starts in the vicinity of the critical point $F$ or $G$, it goes towards $z=1$ plane in a spiral-like manner (see figure \ref{FG}), and the values of cosmological parameters oscillate.  The nearer the trajectory goes towards of  $z=1$ plane the amplitude of oscillation decreases and the values of deceleration parameter oscillate between $-0.18$ to $-0.38$ (approx) whereas  the values of effective equation of state parameter oscillate  between $-0.4$ to $-0.6$ (approx).   On the other hand, the density parameters oscillate between $0.4$ to $0.6$ (a scaling solution) near the $z=1$ plane (see figure \ref{Param}). But, immediately after, the trajectory touches $z=1$ plane the amplitudes of oscillation of the cosmological parameters increase rapidly as the critical points $F$ and $G$ are unstable focus in nature.  Thus, the strange cosmological behavior is to be observed in the vicinity of $z=1$ plane.  The nature of solutions bifurcates from scalar field dominated to matter dominated by passing through the scaling solution and deceleration to acceleration era of expansion by passing through the quintessence boundary and the other way round.  So, in our model, $z=1$ plane turns out to be a strange attractor in the perspective of strange cosmological behavior corresponding to the vector field near the plane.

We have also investigated the evolution of cosmographical parameters with the help of statefinder pair diagnostic tool and examined the reliability of our model by comparing with the observational data such as Hubble parameter and SNIa. From the statefinder diagnostic analysis, it has been found that the model explains the late-time cosmic acceleration and also transits from early decelerating phase to an accelerating phase. It is also observed that the model shows proximity with observational data very well.
The tachyon model presents a compelling alternative to conventional cosmological scenarios, offering richer and more complex dynamics. Its potential to describe certain aspects of the universe warrants further attention. A thorough statistical evaluation, using data from Type Ia Supernovae, Baryon Acoustic Oscillations, the Cosmic Microwave Background, and measurements of large-scale structure (such as $\sigma_8$), is essential to test its viability and refine its parameters. While this paper lays the groundwork, such comprehensive analysis is beyond its current sope and is reserved for future research efforts.

\section*{Acknowledgments}
The authors would like to thank the anonymous referees for their critical review and suggestions as a result of which the paper has improved significantly. The author Goutam Mandal acknowledges UGC, Government of India for providing Senior Research Fellowship [Award Letter No. F.82-1/2018(SA-III)] for Ph.D. The authors would like to thank Md. Arif Shaikh for helpful discussions. The authors are also grateful to Dr. Supriya Pan for helpful discussion on interaction in dark sectors. SKB would like to acknowledge Inter-University Centre for Astronomy and Astrophysics (IUCAA), India for providing research facilities.

\newpage

\end{document}